\documentclass[prodmode,acmtomm,dvipsnames]{acmsmall} 

\makeatletter 
\newcommand\subparagraph{%
  \@startsection{subparagraph}{5}
  {\parindent}
  {3.25ex \@plus 1ex \@minus .2ex}
  {-1em}
  {\normalfont\normalsize\bfseries}}
\makeatother

\usepackage{epsfig}
\usepackage{amsmath}
\usepackage{amssymb}
\usepackage{graphics}
\usepackage{epstopdf}
\usepackage{color}
\usepackage{xcolor}
\usepackage{multirow}
\usepackage{caption}
\usepackage{cite}
\usepackage{array}
\usepackage{url}
\usepackage{subfig}
\usepackage{bibspacing}
\usepackage{slashbox}
\usepackage{placeins}
\usepackage{titlesec}
\usepackage{helvet}
\usepackage{bm}
\usepackage[normalem]{ulem}
\usepackage{mathtools}
\DeclarePairedDelimiter\ceil{\lceil}{\rceil}
\DeclarePairedDelimiter\floor{\lfloor}{\rfloor}
\DeclareMathOperator*{\argmin}{arg\,min}
\usepackage[ruled]{algorithm2e}
 \newcommand{\ignore}[1]{}
\newboolean{showcomments}
\setboolean{showcomments}{true}
\newcommand{\addref}[0]{\ifthenelse{\boolean{showcomments}} {
    \textcolor{green}{(Add ref(s))}}{}}

\newcommand{\jk}[1]{  \ifthenelse{\boolean{showcomments}}	
{ \textcolor{red}{(JK says:  #1)}} {}  }
\newcommand{\vg}[1]{  \ifthenelse{\boolean{showcomments}}
{ \textcolor{green}{(VG says:  #1)}} {}  }

\titleformat{\subsubsection}[runin]
  {\normalfont\fontsize{9}{11}\sffamily\bfseries\slshape} 
  {\thesubsubsection}
  {1em}
  {}
  
  \titleformat{\paragraph}[runin]
  {\normalfont\fontsize{9}{11}\sffamily\bfseries\slshape} 
  {\thesubsubsection}
  {1em}
  {}

\SetAlFnt{\small}
\SetAlCapFnt{\small}
\SetAlCapNameFnt{\small}
\SetAlCapHSkip{0pt}
\IncMargin{-\parindent}

\begin{document}

\markboth{V. Gokhale et al.}{On QoS-Compliant Telehaptic Communication over Shared Networks}

\title{On QoS-Compliant Telehaptic Communication over Shared Networks}
\author{Vineet Gokhale
\affil{Indian Institute of Technology Bombay, University of South Bohemia}
Jayakrishnan Nair
\affil{Indian Institute of Technology Bombay}
Subhasis Chaudhuri
\affil{Indian Institute of Technology Bombay}
Jan Fesl
\affil{University of South Bohemia}
}

\begin{abstract}
  The development of communication protocols for teleoperation with
  force feedback (generally known as \textit{telehaptics}) has gained widespread interest over the
  past decade. Several protocols have been proposed for performing
  telehaptic interaction over shared networks. However, a comprehensive
  analysis of the impact of network cross-traffic
  on telehaptic streams, and the feasibility of Quality of Service
  (QoS) compliance is lacking in the literature. In this paper, we
  seek to fill this gap. Specifically, we explore the QoS experienced
  by two classes of telehaptic protocols on shared networks ---
  Constant Bitrate (CBR) protocols and adaptive sampling based
  protocols, accounting for CBR as well as TCP cross-traffic. Our
  treatment of CBR-based telehaptic protocols is based on a
  micro-analysis of the interplay between TCP and CBR flows on a
  shared bottleneck link, which is broadly applicable for performance
  evaluation of CBR-based media streaming applications. Based on our
  analytical characterization of telehaptic QoS, and via extensive
  simulations and real network experiments, we formulate a set of sufficient conditions for
  telehaptic QoS-compliance. These conditions provide guidelines for
  designers of telehaptic protocols, and for network administrators to
  configure their networks for guaranteeing QoS-compliant telehaptic
  communication.
\end{abstract}





\begin{bottomstuff}
  A preliminary version of this article was presented at IEEE International Symposium on Haptic Audio-Visual Environments and Games (HAVE), 2017.\\
  \noindent Authors' address: V. Gokhale, J. Nair and S. Chaudhuri,
  Department of Electrical Engineering, Indian Institute of Technology
  Bombay, Mumbai 400076, India; emails: \{vineet, jayakrishnan.nair,
  sc\}@ee.iitb.ac.in; V. Gokhale and J. Fesl, Institute of Applied
  Informatics, University of South Bohemia, \v Cesk\' e Bud\v ejovice 37005, Czech
  Republic; emails: \{vgokhale, jfesl\}@prf.jcu.cz.
\end{bottomstuff}

\maketitle
\thispagestyle{empty}
\pagestyle{empty}


\section{INTRODUCTION}
\label{sec:intro}
The past two decades have witnessed rapid advancements in the science
of exploration and manipulation of remote objects with the
augmentation of force feedback -- a field generally referred to as
\emph{telehaptics}. The primary aim of telehaptics is to provide a
touch-based immersive environment to the human user for efficiently
controlling a remote object. Typically, this necessitates
ultra low latency transmission of haptic, auditory and visual
information over a communication network.
Specifically, for a seamless telehaptic interaction, stringent Quality
of Service (QoS) constraints need to be met for each media
type. Table~\ref{table:qos} summarizes the QoS requirements for
telehaptic communication in terms of three important metrics: frame delay,
jitter, and packet loss \cite{ref:qos}.
\begin{table}[h]
\centering
\resizebox{6.6cm}{0.7cm}{
\begin{tabular}{|c|c|c|c|}%
      \hline
       \textbf{Media} & Delay (ms) & Jitter (ms) & Loss (\%) \\ \hline
       Haptic & 30 & 10 & 10 \\ \hline
       Audio & 150 & 30 & 1 \\ \hline
       Video & 400 & 30 & 1 \\ \hline
\end{tabular}
}
\caption{QoS specifications for frame delay, jitter, and packet loss for a smooth telehaptic communication.}
\label{table:qos}
\end{table}

In general, non-conformance to the above QoS constraints results in a
loss of synchronization between the human operator and the remote
environment, resulting
in 
a degraded perception of the remote environment. Specifically,
violating the haptic QoS constraints destabilizes the global haptic control
loop leading to catastrophic effects on the application. Thus, QoS
compliance plays a crucial role in achieving a smooth telehaptic activity.

It is typically infeasible to deploy 
dedicated networks for the purpose of teleoperation.
Moreover, the ubiquitous Internet practically connects every remote
corner of the world. Therefore, it is pragmatic to utilize the
existing networking resources for teleoperation rather than relying on
dedicated resources.  However, the internet, or any shared network, is
utilized simultaneously by several traffic flows. As a result, the
overall cross-traffic seen by the telehaptic application is both
unknown as well as time-varying. This makes telehaptic QoS compliance
on shared networks extremely challenging.

Several protocols have been designed specifically for telehaptic
communication on shared networks
\cite{ref:packetizationintfujimoto,ref:alphan,ref:admux,ref:vhmux,ref:vineetncc,ref:dpm}. However,
the performance evaluation of these protocols has only been carried
out in highly controlled and simplistic network settings. For example,
typically, either no cross-traffic or only constant bit rate (CBR)
cross-traffic is considered in the evaluation of these
protocols. However, in real-world networks, a majority of the traffic
is comprised of Transmission Control Protocol (TCP) flows
\cite{ref:90percentyao,ref:90percentryu}, which are rate-adaptive in
nature. Thus, the evaluation of any telehaptic protocol is incomplete
without analyzing its interplay with TCP cross-traffic.

In this paper, we provide a comprehensive assessment of the interplay
between telehaptic traffic and heterogeneous cross-traffic, consisting
of CBR as well as TCP flows.
This leads to the formulation of a set of sufficiency conditions for
telehaptic QoS compliance. Our analysis is focused on the following two
classes of telehaptic protocols.
%
%
%
\begin{enumerate}
\item {\bf CBR-based telehaptic protocols:} This class of protocols
  generates a constant bitrate (CBR) data stream, i.e., they inject
  traffic into the network at a steady rate. Examples of such
  protocols include the Application Layer Protocol for HAptic
  Networking (ALPHAN) \cite{ref:alphan}, Adaptive Multiplexer (AdMux)
  \cite{ref:admux}, Haptics over Internet Protocol (HoIP)
  \cite{ref:vineetncc}, and the protocol proposed in
  \cite{ref:packetizationintfujimoto}.  Interestingly, a recently
  proposed delay-based rate adaptive protocol \cite{ref:dpm} also
  generates a CBR data stream in presence of TCP traffic. Hence, under
  TCP cross-traffic conditions the rate-adaptive protocol in
  \cite{ref:dpm} also belongs to the class of CBR-based protocols.
\item {\bf Adaptive sampling based telehaptic protocols:} This class
  of protocols employs the adaptive sampling scheme to compress the
  haptic signal
  \cite{ref:telepresence,ref:perception,ref:exploring,ref:levelcrossing}.
  The idea behind adaptive sampling strategy is to identify \emph{perceptually
    significant} haptic samples; transmitting only these samples leads
  to a substantial reduction in long-term average telehaptic data
  rate. Several papers propose telehaptic communication using adaptive
  sampling
  \cite{ref:communication,ref:pahcp,ref:vhmux,ref:opportunistic}.
\end{enumerate}



\ignore{ A few protocols are rate adaptive in nature irrespective of
  the type of cross-traffic.  One such protocol is called
  \emph{visual-haptic multiplexing} \cite{ref:vhmux} for transmission
  of haptic and video data based on \emph{Weber sampler} - a haptic
  data compression scheme.  The Weber sampler identifies the
  perceptually significant force samples that are potentially
  perceivable by the human user, and hence are the ones to be
  transmitted.  Based on the occurrence of such samples the protocol
  in \cite{ref:vhmux} schedules the transmission of video
  frames. Unlike the previous protocols, the instantaneous data rate
  depends heavily on the nature of the generated force signal. Hence,
  the telehaptic stream in this case is variable bit rate (VBR) in
  nature.}  

For the above two classes of protocols, we investigate the interplay
between telehaptic stream and heterogeneous cross-traffic consisting
of TCP and CBR flows. Our contributions are the following.
\begin{enumerate}
\item We develop a mathematical model for analyzing the interplay
  between TCP and CBR flows sharing a single botteneck link, resulting
  in an analytical characterization of delay and jitter
  experienced by the CBR flow (see Section~\ref{sec:tcp}). This
  methodology can be used for performance evaluation of any CBR-based
  streaming protocol in the presence of heterogenous (TCP and CBR)
  cross-traffic.
\item We utilize the above framework to characterize delay and
  jitter experienced by CBR based telehaptic protocols in the presence
  of TCP and CBR cross-traffic (see Section~\ref{sec:cbrtele}). We
  validate these characterizations through simulations and network
  experiments, and subsequently formulate a set of sufficiency
  conditions for telehaptic QoS compliance for CBR based telehaptic
  protocols on shared networks (see
  Section~\ref{sec:validation}). Finally, we show that meeting the
  haptic delay constraint implies meeting the delay constraint for
  audio and video under reasonable media multiplexing mechanisms (see
  Section~\ref{sec:compliance}).
\item For adaptive sampling based protocols, we perform a
  simulation-driven study to show that the statistical compression
  provided by the adaptive sampling strategy provides no meaningful
  economies in terms of network bandwidth requirement. Further, we
  consider the multiplexing protocol in \cite{ref:vhmux} as a working
  example, and demonstrate that uneven packet sizes can result in QoS
  violations on the packet loss criteria. Finally, we provide some important
  guidelines crucial for the design of telehaptic communication protocols that are
  based on adaptive sampling scheme (see Section~\ref{sec:tcpweber}).
\end{enumerate}

\subsection{Typical Telehaptic Environment}
\label{subsec:telehapenvt}
\begin{figure}[!h]
\begin{minipage}[b]{0.4\linewidth}
\centering
\includegraphics[height=2.6 cm,width = 5.5 cm]{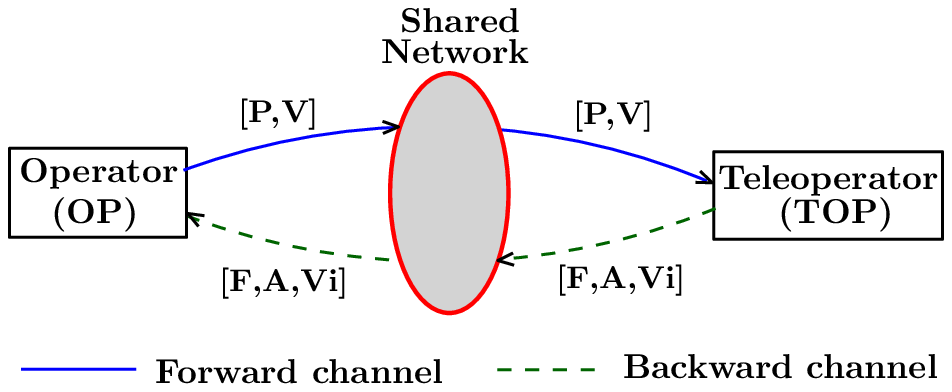}
\caption{Schematic of a point-to-point telehaptic communication framework. Notations: [P, V] - [position, velocity],
[F, A, Vi] - [force, audio, video].}
\label{fig:telehaptic}
\end{minipage}
 \hspace{0.25cm}
\begin{minipage}[b]{0.55\linewidth}
\centering
\includegraphics[height=2.6 cm,width = 8 cm]{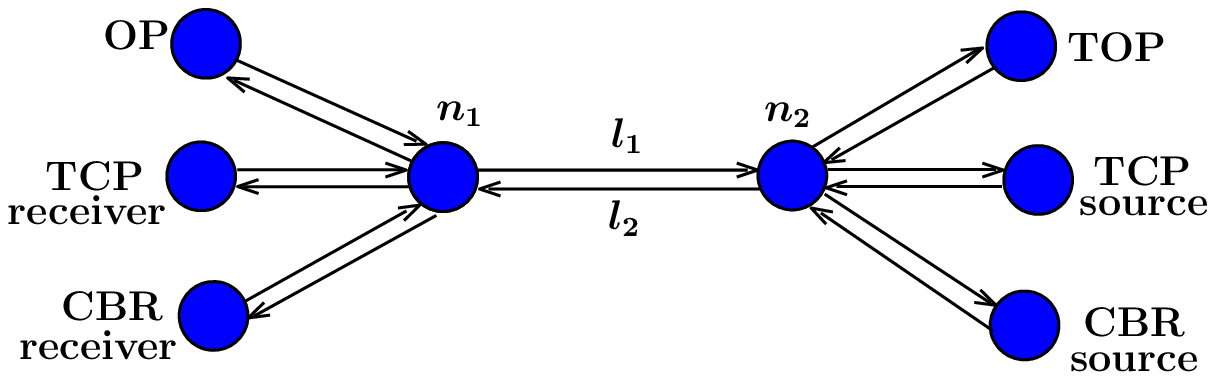} 
\caption{Single bottleneck network topology showing the telehaptic and cross-traffic sources and receivers. $l_1$ and $l_2$ -  bottleneck links on forward and backward channels;
$n_1$ and $n_2$ - intermediate nodes.}
\label{fig:topology}
\end{minipage}
\end{figure}


We now describe the framework of a typical point-to-point telehaptic
communication system on a shared network, shown in
Figure~\ref{fig:telehaptic}. The human operator (OP) controls the
remote robotic manipulator known as the teleoperator (TOP). The OP
transmits the current position and velocity commands on the
\emph{forward channel}. The TOP follows the trajectory of the OP
through execution of the received commands. The resulting force
feedback is transmitted back to the OP along with the captured audio
and video signals on the \emph{backward channel}. Note that the telehaptic
communication is inherently bidirectional and asymmetric in nature.

\subsection{Organization of the Article}
\label{subsec:organization}
This article is organized as follows. In Section~\ref{sec:tcp}, we
provide a brief overview of the working principle of TCP, and present
our proposed mathematical model for characterizing the interplay
between TCP and CBR flows.  In Section~\ref{sec:cbrtele}, we
specialize our model for characterizing the QoS parameters for CBR-based
telehaptic protocols. We validate our claims through rigorous
simulations and network experiments in
Section~\ref{sec:validation}. In Section~\ref{sec:tcpweber}, we
present our results on the interplay between a telehaptic
communication protocol employing the adaptive sampling scheme and
network cross-traffic. We address audio and video delays for CBR-based
telehaptic protocols in Section~\ref{sec:compliance}. Finally, we
review the related literature on the interplay between TCP and CBR traffic in
Section~\ref{subsec:relatedwork}, and conclude in
Section~\ref{sec:conclusions}.

\section{TCP-CBR Interplay}
\label{sec:tcp}

TCP forms the backbone of a wide range of internet applications that
demand reliable data transfer, such as web browsing, email, file
download, and even video streaming applications like YouTube and
Netflix.  Studies show that TCP traffic constitutes over 90\% of all
internet traffic \cite{ref:90percentryu,ref:90percentyao}. TCP is a
transport layer protocol that controls the rate at which the
application injects traffic into the network based on the perceived
network conditions. It achieves end-to-end reliability through
retransmission of lost packets, which are detected using packet
acknowledgments (ACKs) that are sent to the source by the receiver. In
this section, we 
provide an analytical characterization of the delay and the jitter
encountered by a CBR stream co-existing with a TCP stream on a single
bottleneck link. This analysis, which generalizes the work of
\cite{ref:tcpbuffer} on queue dynamics of a single TCP flow, is of
independent interest, shedding light on the interplay between TCP and
CBR streams in a network. Further, our results can be applied to
analyze the performance of CBR-based steaming media applications on
shared networks. In Section~\ref{sec:cbrtele}, we apply these results
to analyze QoS compliance of CBR-based telehaptic protocols that
coexist with TCP cross-traffic on a shared network.

We begin by providing a brief overview of TCP NewReno
\cite{ref:newreno}, which is the most widely deployed variant of TCP
on the internet.

\subsection{TCP Background}
\label{subsec:tcpbg}

A TCP source maintains a variable called \emph{congestion window}
(denoted by $W$) that defines the number of TCP packets that are
\textit{outstanding}, i.e., transmitted but not yet acknowledged. The
congestion window $W$ controls the rate at which TCP traffic is injected
into the network -- a higher $W$ corresponds to a higher transmission
rate, and vice-versa. The TCP source increments $W$ by 1 every round
trip time (RTT). This phase is commonly referred to as
\emph{congestion avoidance} in the literature. Once a packet loss is
detected, TCP infers that the network is overloaded and cuts its
transmission rate aggressively. This phase is referred to as
\emph{fast retransmit, fast recovery} in the literature, wherein the
TCP source retransmits the lost packet and awaits the corresponding
ACK.  Once this ACK is received, the TCP source re-enters the
congestion avoidance phase with an initial congestion window that is
half the window size at the time the loss was detected.\footnote{This
  description assumes a single packet loss; the congestion window
  dynamics are more complicated if there are multiple losses
  \cite{ref:newreno}.}

\begin{figure}[!t]
\centering
  \subfloat[]{\includegraphics[height = 45mm, width = 73mm]{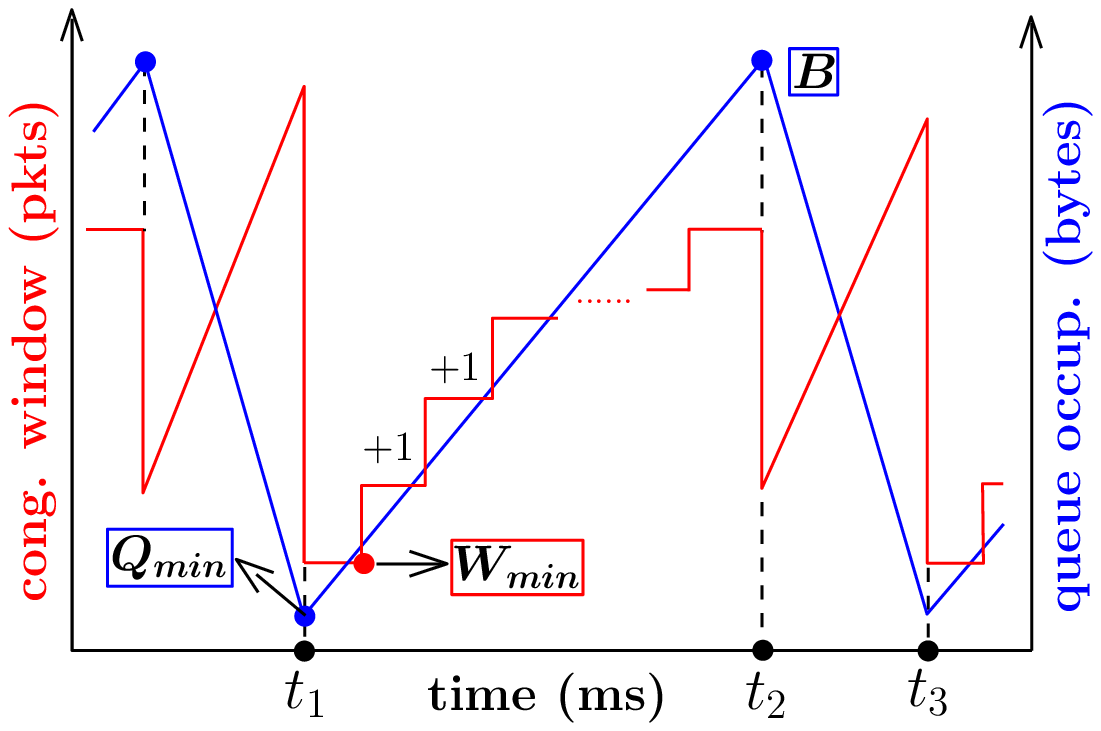} \label{fig:cwnd}}
  \subfloat[]{\includegraphics[height = 45mm, width = 73mm]{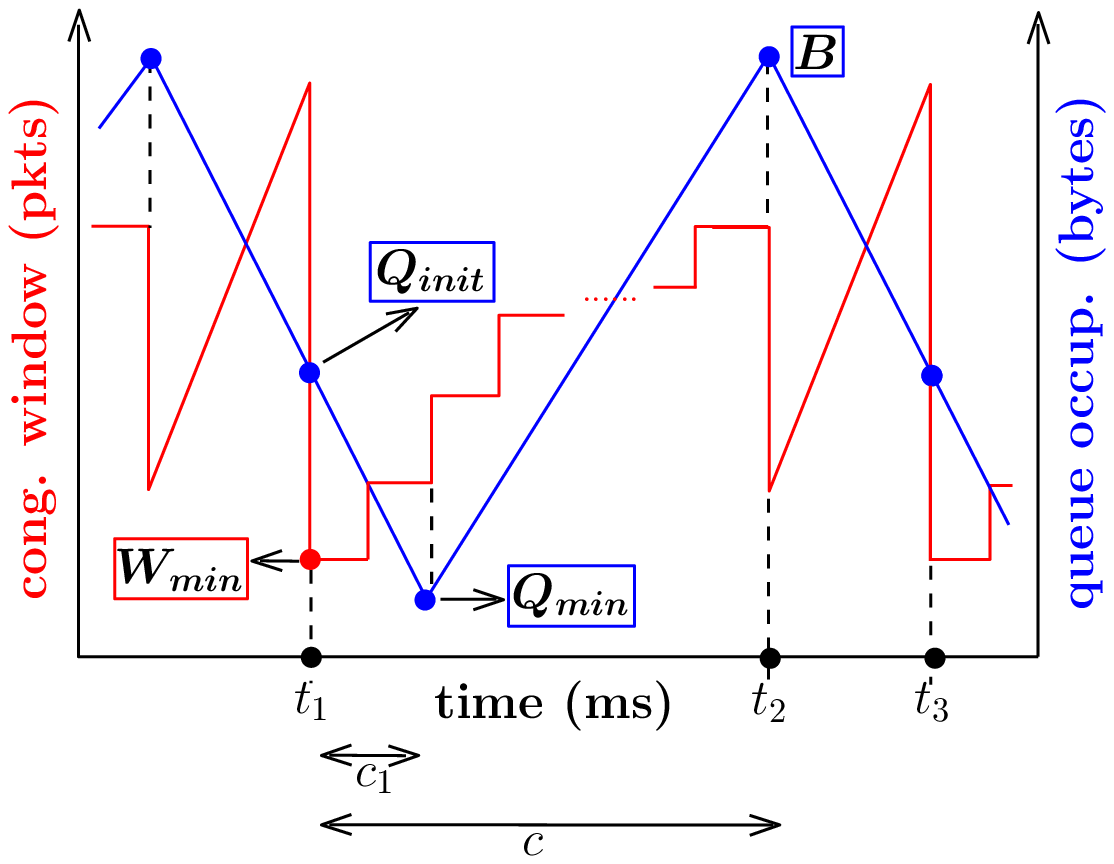} \label{fig:cwndqueue}}
  \caption{Evolution of TCP congestion window and bottleneck queue
    occupancy for (a) single TCP flow (b) heterogeneous flows involving
    single TCP flow and at least one CBR flow.}
\label{fig:tcpcwndbuffer}
\end{figure}

To provide a concrete visualization of the rate adaptation, consider the single bottleneck
network topology shown in Figure~\ref{fig:topology} with a single TCP
source (we ignore telehaptic and CBR cross-traffic for now).
Let $\mu$ denote the capacity (in kbps) of the bottleneck link $l_2$,
and $B$ denote the queue size (in bytes) at the ingress of the
bottleneck link ($n_2$). Let $\tau$ (in ms) denote the one-way propagation
delay of the TCP flow.  The reference \cite{ref:tcpbuffer}
demonstrates that in such a setting, the congestion window $W$ and the
queue occupancy $Q$ on the bottleneck link exhibit a cyclic (periodic)
variation, as shown in Figure~\ref{fig:cwnd}.  The interval between
$t_1$ and $t_2$ corresponds to the congestion avoidance phase. Note
that $W$ is incremented in steps of 1, and the resulting increase in
transmission rate causes the queue occupancy to increase. The
duration between two consecutive updates in $W$ during the congestion
avoidance is termed as a \textit{slot} \cite{ref:tcpbuffer}.  Once a
packet loss (due to queue overflow) is detected (at $t_2$), the source
enters the fast retransmit, fast recovery phase. In this phase,
corresponding to the interval between $t_2$ and $t_3$, the source
retransmits the lost packet, and reduces its transmission rate
aggressively causing the queue to drain quickly.\footnote{Note that
  during fast retransmit, fast recovery phase the source increases $W$
  by 1 for every ACK received. However, no packets are transmitted
  until the time $W$ is less than its value at the time of
  loss. Hence, $W$ does not represent the number of outstanding
  packets during this phase.} Once the source receives the ACK
corresponding to the retransmitted packet (at $t_3$), it re-enters
the congestion avoidance phase, and the cycle repeats.

During the congestion avoidance phase, note that when an ACK arrives
after the start of a slot
the source transmits a packet to compensate for the packet that has
left the network. On the other hand, at the start of a slot (when the
congestion window is increased by 1) the source transmits two
packets back-to-back. While the first one is the \textit{compensating packet},
the second one adds an extra packet into the network for satisfying
the updated $W$ value. We term this additional packet as the
\textit{probing packet}, since this packet probes the network for
extra bandwidth. In other words, the source transmits a probing packet
at the start of each slot. It is worth noting that the length of a
slot is simply the RTT encountered by the corresponding probing
packet.

Let $W_{min}$ and $Q_{min}$ denote the minimum value of $W$ and $Q$ in
one cycle, respectively. The reference \cite{ref:tcpbuffer} provides an
analytical characterization of $W_{min}$ and $Q_{min}.$ Specifically, it is
proved that if $B > 2\mu\tau,$
 \begin{equation*}
    Q_{min} = \frac{B-2\mu\tau}{2}, \qquad W_{min} = \frac{B+2\mu\tau}{2S_{tcp}},
\end{equation*}
where $S_{tcp}$ is the size of a TCP packet.

In the following, we generalize the analysis in~\cite{ref:tcpbuffer}
to include a CBR flow co-existing with the TCP flow on the bottleneck
link. This non-trivial generalization leads to a characterization of
(i)~the maximum and minimum end-to-end CBR delay, (ii)~the maximum CBR
jitter. This characterization will be useful in our subsequent
analysis of the interplay between heterogeneous cross-traffic and
telehaptic stream.


\subsection{TCP-CBR Interplay}
\label{sec:TCP_analysis}


For our analysis, we consider the same network setting as above, except that
there are now two traffic flows on the network, a TCP flow and a CBR
flow.
Let $R$ denote the data rate of the CBR flow.
For simplicity, we assume that the reverse channel (i.e., link $l_1$)
is uncongested.\\

\noindent\textbf{Assumptions:} For the ease of our analysis, we follow
\cite{ref:tcpbuffer} and make the below assumptions.
\begin{enumerate}
\item The TCP source has an infinite backlog of data.
\item The access links to $l_1$ and $l_2$ have very high bandwidth and
  negligible propagation delays.  In effect, the traffic sources
  (respectively, receivers) directly feed into (respectively, read
  from) $n_2$ (respectively, $n_1$).
\item The queue size at the ingress of the bottleneck link ($n_2$) is
  greater than the bandwidth-delay product of the TCP flow i.e.,
  $B > 2\mu\tau$. This condition guarantees that the queue never
  empties, and hence the bottleneck link is never
  underutilized~\cite{ref:tcpfull}.
\item In every cycle, the TCP stream loses exactly one packet due to
  queue overflow at~$n_2$.
\end{enumerate}

\subsubsection{Characterization of Queue Occupancy}
\label{subsubsec:queue}




Similar to the case of a single TCP flow in \cite{ref:tcpbuffer}), it
can be shown that in steady state $W$ and $Q$ vary periodically in
time; see Figure~\ref{fig:cwndqueue}. However, the presence of the CBR
traffic changes the nature of the queue occupancy evolution
relative to the congestion window evolution. Note that during the
congestion avoidance phase (the interval between $t_1$ and $t_2$), the
queue occupancy initially decreases over $c_1$ slots, and then
increases until an overflow occurs. Let $c$ denote the total number of
slots in the congestion avoidance phase of each cycle.

Let $Q_{init}$ denote the queue occupancy at the start of the
congestion avoidance phase.  Let $i$ be the slot index, and $Q(i)$
denote the value of $Q$ at the start of $i$\textsuperscript{th}
slot. Therefore, we can write $Q(1) = Q_{init}$.  For brevity, we only
present the key results of our analysis in this section. Interested
readers can refer to Appendix~\ref{append:queue} for a detailed
description.  From our analysis, we obtain closed form expressions for
$W_{min}, Q_{init}$, and $c$ as given by Equations \eqref{equ:wmin},
\eqref{equ:qinit}, and \eqref{equ:c}, respectively.
\begin{equation}
\label{equ:wmin}
W_{min} = \frac{(B+2\mu\tau)(1-\alpha)}{2S_{tcp}},
\end{equation}
\begin{equation}
\label{equ:qinit}
Q_{init} = \frac{(B+2\mu\tau)(1+\alpha)}{2}-2\mu\tau,
\end{equation}
\begin{equation}
\label{equ:c}
c = \frac{(B+2\mu\tau)(1-\alpha)}{2S_{tcp}} + 1,
\end{equation}
where $\alpha := \frac{R}{\mu}$. Additionally, the analysis also
yields Equations~\eqref{equ:qmin1} and \eqref{equ:qmin2} with unknowns
$Q_{min}$ and $c_1$.
\begin{equation}
\label{equ:qmin1}
Q_{min}+\Big[\frac{(B+2\mu\tau)(1-\alpha)}{2 S_{tcp}}-c_1+1\Big]\Big[\frac{S_{tcp}}{1-\alpha}\Big] = B
\end{equation}
\begin{equation}
\label{equ:qmin2}
Q_{min} = \Big[\frac{(B+2\mu\tau)(1+\alpha)}{2}-2\mu\tau\Big]\alpha^{c_1} + \Big[\frac{(B-2\mu\tau)(1-\alpha^{c_1})}{2}\Big] +  S_{tcp} \sum_{j = 0}^{c_1-2} (c_1-1-j)\alpha^j
\end{equation}

As can be noticed, $Q_{min}$ and $c_1$ do not admit a closed form
characterization. However, they are easily amenable to numerical
computation as follows.
The evolution of the queue occupancy across slots during the congestion
avoidance phase is given by:
\begin{equation}
 \begin{split}
   Q(i) = \Big[\frac{(B+2\mu\tau)(1+\alpha)}{2}-2\mu\tau\Big]\alpha^{i-1} + \Big[\frac{(B-2\mu\tau)(1-\alpha^{i-1})}{2}\Big]
+ S_{tcp} \sum_{j = 0}^{i-3} (i-2-j)\alpha^{j}, \\ \forall i \in [1,c].
\end{split}
\label{equ:qi}
\end{equation}
Note that \eqref{equ:qmin2} is a special case of
\eqref{equ:qi} setting $i = c_1 + 1$ (which corresponds to the queue
occupancy after $c_1$ slots).
Since $Q(\cdot)$ is unimodal over a cycle, $Q_{min}$ can be computed
numerically by minimizing $Q(i)$ over $i \in [1,c].$ Therefore, we have
\begin{equation}
\label{equ:qminc1}
Q_{min} = \min_{i\in [1,c]} Q(i), \quad  c_1 = \argmin_{i\in [1,c]} Q(i)-1. 
\end{equation}

As described previously, TCP rate adaptation is based on queue
overflows which occur when the queue occupancy reaches the maximum
permissible value $B$. This implies that the maximum queue occupancy
$Q_{max} = B$. Therefore, we can write $Q(i) \in [Q_{min}, Q_{max}]$.

To summarize, the above analysis characterizes the minimum and the
maximum queue occupancy at the ingress of the bottleneck link in terms of network
parameters ($\mu, \tau, B$) and the CBR source parameter $R$.

\subsubsection{Characterization of CBR Delay}
\label{subsubsec:delayChar}
Based on the above results, we now move to the characterization
of the end-to-end delay experienced by the CBR flow.
The delay experienced by the CBR packets is composed of propagation
delay and queueing delay. The latter is in turn proportional to the queue occupancy
encountered by the CBR packet upon arrival into the queue at the
bottleneck link. Thus, the minimum and the maximum end-to-end delay seen
by CBR packets, denoted $d_{min}$ and $d_{max},$ respectively, are
given by
\begin{align}
  \label{equ:dmin}
  d_{min} &= \tau+\frac{Q_{min}}{\mu}, \\
  \label{equ:dmax}  
  d_{max} &= \tau+\frac{B}{\mu}.
\end{align}

To summarize, the CBR delays vary cyclically (in synchronization with
the queue occupancy) over the range [$d_{min}, d_{max}$]. We apply
this delay characterization to determine the telehaptic delays
(Section~\ref{subsec:hapdelay}), and subsequently to derive
sufficiency conditions for QoS compliance of telehaptic flows
(Sections~\ref{sec:validation} and \ref{sec:tcpweber}).


\subsubsection{Characterization of CBR Jitter}
\label{subsubsec:jitterChar}
We now move to characterizing the jitter experienced by the CBR flow
in presence of TCP cross-traffic. Jitter refers to the variation in
the inter-packet delay. Formally, we define the CBR jitter
as $$\delta_{cbr} = \Delta D_j,$$ where $\Delta D_i$ refers
to the difference between the end-to-end delays experienced by
$j$\textsuperscript{th} and $(j-1)$\textsuperscript{th} CBR packets \cite{ref:jitter}.
Therefore, the maximum CBR jitter is given as $$\delta_{cbr(max)} = \max_{j} \Delta D_j.$$
Note that our definition of maximum jitter captures the largest
  \emph{positive} difference between the end-to-end delays experienced
  by successive CBR packets. Indeed, for streaming applications, it is
  these positive delay differences that are troublesome; a negative
  delay difference only means that a sample arrived earlier than its nominal
  rendering time.

Since the only variable component of the end-to-end delay is the
queueing delay, it follows that $$\delta_{cbr(max)} = \max_{j}
\frac{1}{\mu} \Delta Q_j,$$ where $\Delta Q_j$ denotes the
difference between the queue occupancy seen by $j$\textsuperscript{th}
and $(j-1)$\textsuperscript{th} CBR packets.
This results in the following characterization of the CBR jitter.
\begin{equation}
  \label{eq:jitter_1}
  \delta_{cbr(max)} = \frac{1}{\mu} \left[ m_{tcp}S_{tcp} + (R - \mu)
    T_{cbr} \right].
\end{equation}
Here, $T_{cbr}$ refers to the interval between the transmission of
successive CBR packets, and $m_{tcp}$ denotes the maximum number of
packets transmitted by the TCP source in an interval of length
$T_{cbr}$. Thus, all that remains is to characterize $m_{tcp}.$

At this stage, it is appropriate to describe the \textit{cumulative
  acknowledgement} principle of TCP NewReno, as this governs the TCP
parameter $m_{tcp}$.  According to this principle, the TCP receiver
transmits an ACK every $n$\textsuperscript{th} packet, where $n \geq
1$, thereby cumulatively acknowledging the reception of $n$ successive
packets.  Note that the description of the working of TCP in
Section~\ref{subsec:tcpbg} corresponds to $n = 1$. In general, a TCP
source transmits a burst of $n$ packets when a (cumulative) ACK that
acknowledges only compensating packets is
received (this burst consists of $n$ compensating packets), and
transmits a burst of $n+1$ packets when a
(cumulative) ACK that acknowledges a probing packet is
received (this burst consists of $n$ compensating packets and a probing
packet).\footnote{The analysis of the queue occupancy dynamics in
  Section~\ref{subsubsec:queue} remains unaffected by the value of $n,$ since that
  analysis only relies on the (coarser) queue occupancy evolution across
  slots.} Note that packets transmitted in the same burst are not
necessarily acknowledged simultaneously. Specifically, if there
are $m < n$ unacknowledged packets at the receiver when a burst of $n$ packets
arrives, then the receiver cumulatively acknowledges the older $m$ packets and the
earliest $n-m$ packets in the current burst. The remaining $m$ packets of the
current burst are cumulatively acknowledged when the next burst arrives.

\begin{figure}[!h]
\centering
\includegraphics[height = 45mm, width = 125mm]{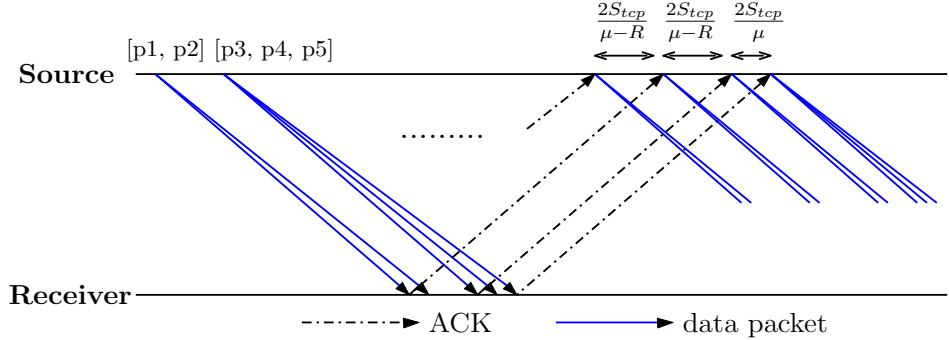}
\caption{Demonstration of the working of cumulative ACK mechanism of TCP with $n =$ 2, along with
the specification of inter-ACK gap. The packets are numbered sequentially in the
order of their transmission, and the packets belonging to the same burst
are grouped together using square brackets.}
\label{fig:cumulack}
\end{figure}

Since the source transmissions are triggered by ACK receptions, $m_{tcp}$
depends on the maximum number of ACKs that can be received by the
source in an interval of length $T_{cbr}.$ Nominally, as we show in
Figure~\ref{fig:cumulack} with $n =$ 2 (explained in detail in Appendix~\ref{append:jitter}),
the interval between successive ACK
receptions at the source equals $\frac{nS_{tcp}}{\mu-R}$, if the ACKs acknowledge
only the probing packets. However, at
a slot boundary, the two ACKs are received with a (smaller) time gap of
$\frac{nS_{tcp}}{\mu}$ due to the following: Suppose that
there are $n-1$ unacknowledged packets at the receiver when the burst of
$n+1$ packets containing a probe packet is received. In this case, the burst would
trigger two ACKs from the receiver, one to acknowledge the $n-1$ older
packets and the first packet of the current burst, and another to
acknowledge the remaining $n$ packets of the current burst. It is
important to note the following.
\begin{itemize}
\item The two ACKs would be received $nS_{tcp}/\mu$ time units
  apart, since this is the time required for $n$ back-to-back packet
  receptions at the receiver.
\item The first ACK would trigger a transmission of $n-$packet burst from
  the source, while the second (which acknowledges a probing packet)
  would trigger a transmission of $(n+1)-$packet burst. 
\end{itemize}

\noindent Thus, $m_{tcp}$ is given by
\begin{equation}
  m_{tcp} = n+1+ n\Bigg(1+\floor*{\frac{T_{cbr}-\frac{nS_{tcp}}{\mu}}{\frac{nS_{tcp}}{\mu-R}}}\Bigg)I_{\big(T_{cbr} > \frac{nS_{tcp}}{\mu}\big)},
  \label{equ:mtcp}
\end{equation}
where $I_z =$ 1 if $z =$ 1, and 0 otherwise. Here, the term $n+1$ is
due to the packets in the $(n+1)-$packet burst. The term inside the
parenthesis represents the maximum number of $n-$packet bursts that
can be transmitted in the interval of length $T_{cbr}$ in addition to
the $(n+1)-$packet burst. In other words, it is the maximum number of
$n-$packet bursts transmitted in an interval of length
$T_{cbr}-\frac{nS_{tcp}}{\mu}$.

From Equations~\eqref{eq:jitter_1} and \eqref{equ:mtcp}, the maximum
jitter experienced by the CBR stream is given as
\begin{equation}
  \delta_{cbr(max)} = \frac{S_{tcp}}{\mu} \Bigg[n+1+n\Bigg(1+\floor*{\frac{T_{cbr}-\frac{nS_{tcp}}{\mu}}{\frac{nS_{tcp}}{\mu-R}}}\Bigg)I_{\big(T_{cbr} > \frac{nS_{tcp}}{\mu}\big)}\Bigg] + \frac{RT_{cbr}}{\mu} - T_{cbr}.
\label{equ:jitter}
\end{equation}

The above characterization allows us to express the maximum jitter of
CBR stream in terms of a set of source parameters ($n, S_{tcp},
T_{cbr}, R$) and a network parameter $\mu$. Interestingly,
$\delta_{cbr(max)}$ has no dependence on other network parameters like $B$
and $\tau,$ both of which influence the delay profile significantly. Therefore,
we conclude that unlike CBR delay, maximum CBR jitter is primarily \textit{source-driven}.

\section{CBR-based Telehaptic Protocols: QoS Characterization}
\label{sec:cbrtele}

In this section, we apply the analytical characterization of TCP-CBR
interplay in Section~\ref{sec:tcp} to analyse the QoS experienced by
CBR-based telehaptic flows in the presence of heterogeneous
cross-traffic. For concreteness in exposition, we assume that the (CBR
based) telehaptic flow encounters one TCP and one CBR cross-traffic
flow over the bottleneck link in the network topology shown in
Figure~\ref{fig:topology}.\footnote{Our results extend easily to the case where
  there are multiple CBR cross-traffic flows, and also multiple synchronized
  TCP cross-traffic flows (as in \cite{ref:tcpbuffer}).} Let $R_h$ and
$R_{cross}$ denote the rates of the telehaptic stream
and the CBR cross-traffic stream, respectively. Note that the aggregate rate of CBR
cross-traffic as seen by the TCP source (using the notation of
Section~\ref{sec:tcp}) equals $R = R_h + R_{cross}.$ Finally, let the
inter-packet gap of telehaptic stream be denoted by $T_h$, and the
packet size of CBR cross-traffic be denoted by $S_{cross}$.

\subsection{Characterization of Haptic Delay}
\label{subsec:hapdelay}

The maximum and the minimum delay experienced by telehaptic packets
follow directly from the analysis in Section~\ref{sec:tcp}. Indeed,
the characterization of the minimum queue occupancy $Q_{min}$ on the
bottleneck link
depends on the aggregate rate $R$ of the CBR cross-traffic seen
by the TCP source (and not on the composition of the CBR cross
traffic). Thus, Equations~\eqref{equ:dmin} and~\eqref{equ:dmax} also
determine the minimum and the maximum delays encountered by the telehaptic
packets, respectively. In other words, haptic delays vary cyclically
in the range $[d_{min}, d_{max}]$.\footnote{Note that we are only
  characterizing the end-to-end delays seen by the \emph{packets}
  generated by the telehaptic stream. The haptic \emph{frames} may
  encounter additional delays depending on the packetization and
  multiplexing mechanism employed. For example, if each telehaptic
  packet contains two haptic frames, then the earlier of these frames
  would experience an additional delay of 1 ms due to packetization.} We
validate these bounds through simulations and real network experiments
in Sections~\ref{subsubsec:delayvalidation} and \ref{subsubsec:delayExpts},
respectively.


\subsection{Characterization of Haptic Jitter}
\label{subsec:hapjitter}

Next, we turn to the characterization of the maximum jitter experienced by the
telehaptic packets. It is important to note that in practice, negative
haptic jitter (caused by decreasing haptic delays) are canceled using
a jitter buffer at the receiver. The jitter buffer delays
the play-out of the received samples such that the rendering
jitter is minimized. Therefore, in this paper, we focus only
on positive haptic jitter, (as discussed in
Section~\ref{subsubsec:jitterChar}), which has the potential to impair
human perception during a telehaptic activity.

Unlike delay, the jitter experienced by the two CBR streams (the
telehaptic stream and the cross-traffic stream with rate $R_{cross)}$)
will in general be different. This is because the jitter of each stream
is determined by the maximum cross-traffic injected into the queue between
successive packets of \emph{that} stream. Accordingly, in the
following, we adapt the jitter characterization in
Section~\ref{subsubsec:jitterChar} to obtain an expression for the maximum
haptic jitter.


Analogous to Equation~\eqref{eq:jitter_1}, the maximum haptic jitter can be expressed as
\begin{equation*}
\delta_{h(max)} = \frac{1}{\mu}[m_{tcp}S_{tcp} + m_{cross}S_{cross} + (R_h - \mu) T_h],
\label{equ:nuh}
\end{equation*}
where $m_{cross}$ denotes the maximum number of CBR cross-traffic
packets transmitted within an interval of length $T_h$. The
numerator in the above expression indicates the maximum increase in the queue occupancy
between the arrival of two successive telehaptic packets. 
Analogous to Equation~\eqref{equ:mtcp}, we
can write the maximum number of TCP packets that can be transmitted in
the interval $T_h$ as
 \begin{equation*}
   m_{tcp} = n+1+n\Bigg(1+\floor*{\frac{T_h-\frac{nS_{tcp}}{\mu}}{\frac{nS_{tcp}}{\mu-R}}}\Bigg)I_{\big(T_h > \frac{nS_{tcp}}{\mu}\big)}.
\label{equ:deltaQh}
\end{equation*}
Moreover, it is easy to see that 
\begin{equation*}
m_{cross} = \ceil*{\frac{R_{cross}T_h}{S_{cross}}}.
\label{equ:mcbr}
\end{equation*}

\noindent Combining the above equations, the maximum haptic jitter is given by
\begin{equation}
  \delta_{h(max)} = \frac{\Bigg[n+1+n\Bigg(1+\floor*{\frac{T_h-\frac{nS_{tcp}}{\mu}}{\frac{nS_{tcp}}{\mu-R}}}\Bigg)I_{\big(T_h > \frac{nS_{tcp}}{\mu}\big)}\Bigg]S_{tcp}+ \ceil*{R_{cross}T_h} + R_hT_h}{\mu} - T_h
\label{equ:jitter}
\end{equation}
We validate Equation~\eqref{equ:jitter} experimentally in
Sections~\ref{subsubsec:jitter} and \ref{subsubsec:jitterExpts}.\footnote{Note that we are characterizing
  \emph{packet-level} jitter here. The \emph{frame-level} haptic
  jitter will also depend on the multiplexing and packetization
  mechanism employed. However, if each telehaptic packet contains a
  single haptic frame, then the packet-level jitter matches the
  frame-level jitter.}

\ignore{
\subsection{Characterization of Packet loss} 
\label{subsec:haploss}
Our characterization of delay and jitter in this paper is based on the
fluid-model approximation of the traffic flows that treats the packet
flows as continuous-time entities. However, packet losses are
discrete-time events, and therefore it is hard to characterize them
using this approximation. In this paper, we instead resort to a
simulation-based characterization of the packet loss (see
Section~\ref{subsubsec:loss}).

In the following section, we test the validity of our characterization
models through extensive simulations using CBR based telehaptic
traffic (Section~\ref{sec:validation}).  For the purpose of
illustration, we consider the telehaptic traffic generated by the
protocol proposed in \cite{ref:vineetncc}. However, it is worth
reemphasizing that our analysis is applicable to the interplay between
TCP and any CBR-based traffic in general. We then move to the
investigation of the interplay between telehaptic traffic generated by
an adaptive sampling scheme and the heterogeneous cross-traffic
(Section~\ref{sec:tcpweber}).  }

\section{CBR-based Telehaptic Protocols: Experimental Results}
\label{sec:validation}

The goal of this section is to validate our analysis
presented in Section~\ref{sec:cbrtele}, and to subsequently develop an
understanding of the conditions required for QoS-compliant telehaptic
communication on a shared network for CBR based telehaptic
protocols.

In the first part of this section, we use NS3 -- a discrete event
network simulator for validating our analytical
model. We find that our delay and jitter bounds are fairly accurate
over a wide range of network settings. We also make the empirical
observation that telehaptic packet losses are rare so long as the
packet sizes are small relative to TCP packets. The above observations
lead us to formulate a comprehensive set of conditions for
QoS-compliant telehaptic communication on shared networks.

To further test the validity of our conclusions under real network
conditions, we also conducted rigorous experiments on a real network. The
results of these experiments are presented in the second part of this
section. The delay and packet loss observations match with those in the
simulations. Interestingly, we find a mismatch between the the measured
jitter in these experiments and our analytical jitter bound. We are
able to trace these errors to differences between the implementation
of TCP NewReno in the employed operating system and the RFC specification. We
conclude that telehaptic jitter is highly sensitive to variations in
the implementation of TCP in the operating systems of the
sender/receiver.

The experimental settings that follow apply to both simulations as well as
network experiments. We employ the single
bottleneck network topology shown in Figure~\ref{fig:topology}. Unless
otherwise specified, we use the following network settings throughout
this section. We set $\mu =$ 6 Mbps, $\tau =$ 8 ms and $B =$ 14
kB.\footnote{The chosen settings represent a medium speed internet
  link of length approximately equal to 1000 miles.}  We work with
real haptic traces generated by the Phantom Omni device
\cite{ref:phantom1} which offers a single point of interaction
between the human user and the haptic environment.
Considering the standard haptic
sampling rate of 1 kHz, and accounting for the overhead due to packet
headers, we get a forward channel data rate $R_f =$ 688 kbps, with
packets of size 86 bytes transmitted every
millisecond~\cite{ref:vineetncc}. On the backward channel, we simulate
audio and video payload at the rate of 64 kbps and 400 kbps,
respectively. We consider the media multiplexing mechanism proposed
in~\cite{ref:vineetncc}, where each packet contains a single haptic
sample and an audio/video fragment of a fixed size. Accounting for the
the packet header overhead leads to a backward channel data rate $R_b =$
1.096 Mbps, with packets of size 137 bytes transmitted every
millisecond.

\ignore{ Each of the force, position and velocity signals is
  represented as a 3-D vector, with each of the components encoded as
  \emph{float} data type (4 bytes).  Considering the standard haptic
  sampling rate of 1 kHz, the haptic payload rates on the forward and
  backward channel amounts to 192 kbps and 96 kbps, respectively.
  Adding a packet overhead of 62 bytes \cite{ref:dpm}, we arrive at
  $R_h =$ 688 kbps on the forward channel. On the backward channel, we
  simulate audio and video payload at the rate of 64 kbps and 400
  kbps, respectively. Adding to this a packet overhead of 67 bytes
  \cite{ref:dpm}, we get $R_h =$ 1.096 Mbps on the backward channel.}

For brevity, we report the results for the case in which cross-traffic
sources are added to the backward channel only. We introduce a TCP
NewReno source with the standard packet size $S_{tcp} =$ 578 bytes. We
also add a CBR cross-traffic source with packet size $S_{cross} =$ 150
bytes.\footnote{This is the typical packet size of a
  video-conferencing application such as Skype.} In the notation of
Section~\ref{sec:TCP_analysis}, note that the aggregate CBR rate on
the backward channel $R = R_b + R_{cross}.$ For sustaining the TCP
flow throughout the duration of the experiment, we need to ensure that
$R < \mu$ so that the TCP flow has sufficient network bandwidth to
perform rate adaptation. Our simulations and network experiments
are performed for a duration
of 500 seconds. Due to the fact that QoS requirements of delay and
jitter for haptic samples are stricter than those for audio/video, we
only report haptic delay and jitter measurements in this section. We
discuss audio/video QoS compliance in
Section~\ref{sec:compliance}. However, we report the packet loss
measurements for all three media types.

\subsection{Simulations}
\label{subsec:simul}
In this section, we present the validation results of our analysis through
simulations.
\subsubsection{Haptic Delay}
\label{subsubsec:delayvalidation}
In our simulations, we observe that the TCP congestion window exhibits
steady behavior only if $R <$~5.5~ Mbps. Hence, we restrict our
measurements to a maximum CBR rate of $R =$~5.5~Mbps, since our
analysis applies only to steady state TCP dynamics.  In
Table~\ref{table:hapDelay}, we report the minimum and the maximum haptic
delays as measured in the simulations and the corresponding analytical
bounds (stated in Section~\ref{subsec:hapdelay}) by varying
$R_{cross}$ to get $R$ in the range [$R_b$, 5.5 Mbps].  Throughout
this range, we see that while the analytical lower bound $d_{min}$ has
a modest accuracy, the upper bound $d_{max}$ is highly accurate.
\begin{table}[h]
\parbox{.5	\linewidth}{
\centering
\resizebox{5.8cm}{1.2cm}{
\begin{tabular}{|c|c|c|c|c|}%
      \hline
      \multirow{2}{*}{$R$ (Mbps)} & \multicolumn{2}{c|}{$d_{min}$ (ms)} & \multicolumn{2}{c|}{$d_{max}$ (ms)}\\
      \cline{2-5}
        & A & S & A & S\\ \hline
       1.096 & 9.91 & 8.89 & 26.66 & 26.47 \\ \hline
       2 & 10.74 & 9.21 & 26.66 & 26.40 \\ \hline
       3 & 12.27 & 11.71 & 26.66 & 26.62 \\ \hline
       4 & 14.81 & 12.95 & 26.66 & 26.45 \\ \hline
       5 & 19.87 & 16.95 & 26.66 & 26.55 \\ \hline
       5.5 & 22.68 & 19.77 & 26.66 & 26.43 \\ \hline
\end{tabular}}
    \captionof{table}{Comparison of $d_{min}$ and $d_{max}$ by analysis (A) and simulation (S) for a wide range of $R$.}
    \label{table:hapDelay}
 }   
\hfill
  \parbox{.45\linewidth}{
  \centering
  \resizebox{3.6cm}{1.2cm}{
  \begin{tabular}{|c|c|c|}
  \hline
  \multirow{2}{*}{$\mu$ (Mbps)}  & \multicolumn{2}{c|}{$\delta_{h(max)}$ (ms)} \\ \cline{2-3}
   & A & S \\ \hline
    9 & 1.46 & 1.46\\\hline
    12 & 1.62 & 1.61\\\hline
    15 & 1.09 & 1.09\\\hline
    18 & 0.84 & 0.85\\\hline
    21 & 0.49 & 0.49\\\hline
    25 & 0.63 & 0.62\\\hline
      \end{tabular}}
      \caption{Comparison of $\delta_{h(max)}$ by analysis (A) and simulation (S) for a wide range of $\mu$.}
      \label{table:jittertable}
      }
\end{table}

\begin{figure}[t]
\begin{minipage}[b]{0.5\linewidth}
\centering
\includegraphics[height=3.8 cm,width = 7.5 cm]{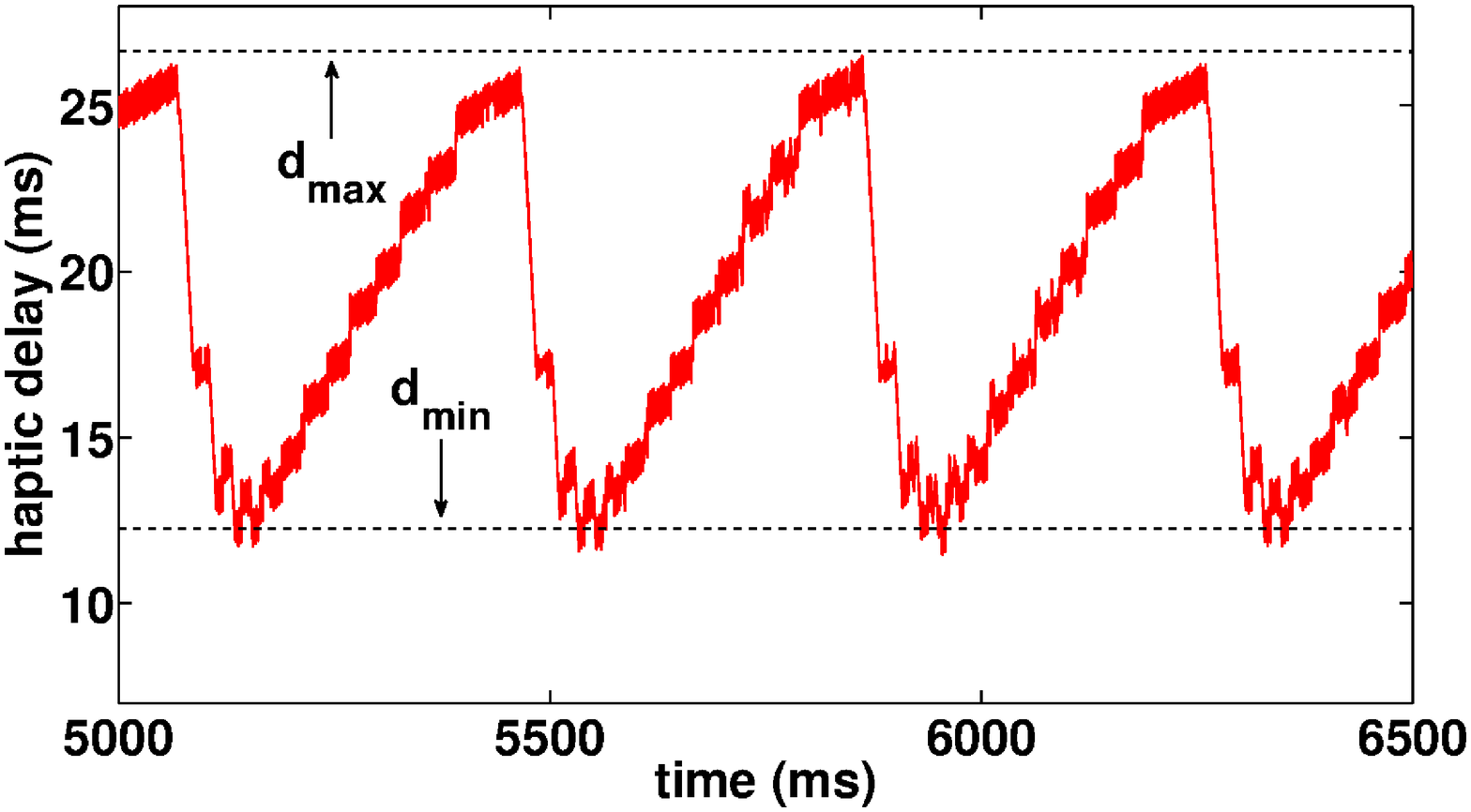}
\caption{Plot of haptic delay showing the corroboration between analytical estimate and simulation measurement.}
\label{fig:hapDelay}
\end{minipage}
\hspace{0.05cm}
\begin{minipage}[b]{0.5\linewidth}
\centering
\includegraphics[height=3.8 cm,width = 7.5 cm]{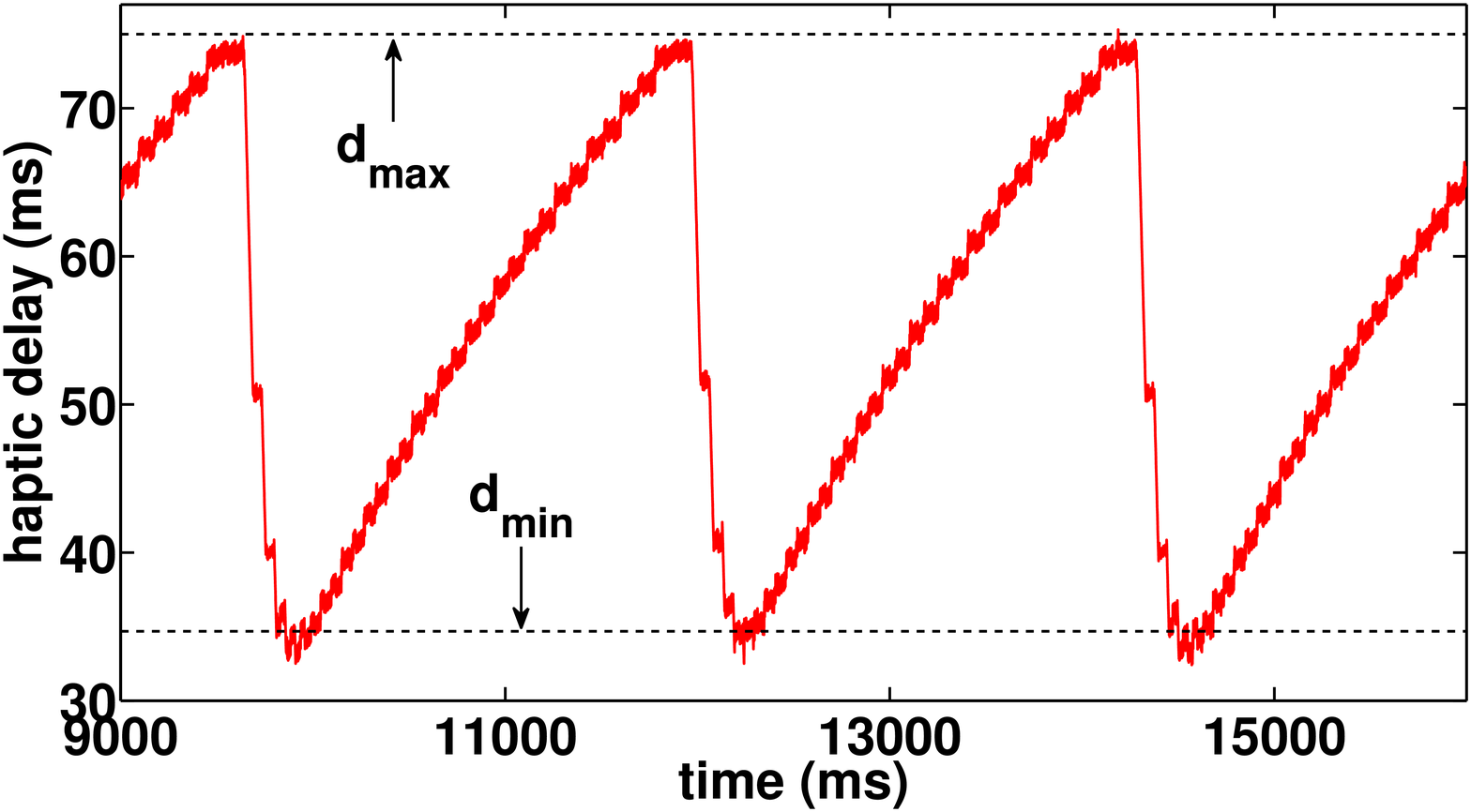} 
\caption{Plot of haptic delay demonstrating the severe QoS violation for a particular network setting.}
\label{fig:hapDelayTypical}
\end{minipage}
\end{figure}

In Figure~\ref{fig:hapDelay}, we plot the temporal variation of the
haptic delay for $R =$ 3 Mbps (i.e., $R_{cross} =$ 1.904 Mbps), along
with the analytical bounds. Note that the haptic delay evolves
periodically over time, matching our analytical upper and lower
bounds.

We make the following remarks.
\begin{itemize}
\item The upper bound $d_{max},$ which is insensitive to $R,$ is
  highly accurate. However, the lower bound $d_{min}$ becomes
  inaccurate as $R$ approaches $\mu$. This is because our
  characterization of $d_{min}$ assumes a single TCP packet loss in
  each cycle; see Assumption (4) in Section~\ref{sec:TCP_analysis}.
  However, we observe in our traces that as $R$ approaches $\mu$, TCP
  starts to lose multiple packets per cycle, leading to a very different
  congestion window evolution from the one analyzed. Simulating a wide
  range of network settings, we observe that a sufficient condition
  for a single TCP packet loss per cycle (and consequently for the
  accuracy of $d_{min}$) is $R \leq 0.65 \mu.$
\item Since the analytical upper bound $d_{max}$ is highly accurate,
  it can be used to check for QoS-compliance of the haptic delay
  for a given network setting, i.e. $d_{max}<$ 30 ms.
  In the network setting under consideration, $d_{max} =$ 26.66 ms
  which is less than the QoS limit of
  30 ms. Indeed, our measurements confirm that the haptic delay
  constraint is satisfied in this case. 
\end{itemize}

To see another example, consider the following setting: $\mu =$ 6
Mbps, $\tau =$ 15 ms, and $B =$ 45 kB. In this case, using \eqref{equ:dmax},
we obtain $d_{max} =$ 75 ms, which suggests that the haptic delay
constraint cannot be met. Indeed, simulations show that this is the
case; see Figure~\ref{fig:hapDelayTypical}. Thus, the expression for
$d_{max}$ can be used to identify the class of network settings where
the QoS-compliance of the haptic delay is feasible.

\ignore{ We observe that beyond $R =$ 3.9 Mbps the error in $d_{min}$
  between the analytical and the simulation measurements increases at
  a faster rate. From the packet traces we discover that beyond this
  point, multiple TCP packets get dropped every time the queue
  overflows. Recall that our analysis applies only to single packet
  loss case. For multiple packet losses, the queue drain time (between
  $t_2$ and $t_3$) increases, leading to a lower $Q_{min}$, and
  thereby a higher error. However, the accuracy of $d_{max}$ is robust
  to $R$. Through simulations over a wide range of $\mu$, we
  empirically identify $R = 0.65 \mu$ as the threshold beyond which
  TCP encounters multiple packet losses per cycle.  }

\begin{figure}[h]
\begin{minipage}[b]{0.5\linewidth}
\centering
\includegraphics[height=3.55 cm,width = 7.5 cm]{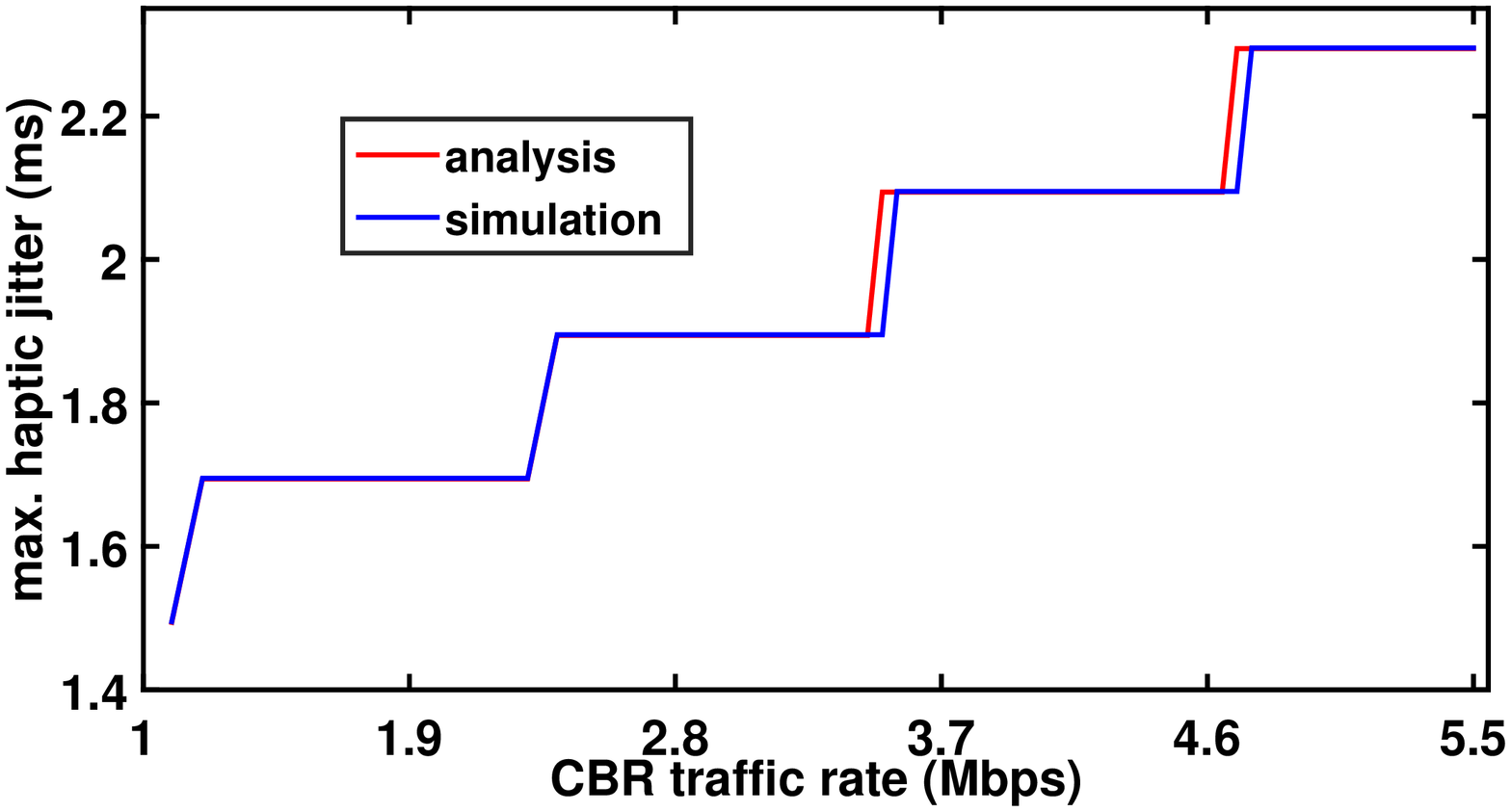} 
\caption{Plot demonstrating the corroboration between theoretical estimates and simulation measurements of maximum haptic jitter.}
\label{fig:jitterPhantom}
\end{minipage}
 \hspace{0.05cm}
\begin{minipage}[b]{0.5\linewidth}
\centering
\includegraphics[height=3.8 cm,width = 7.5 cm]{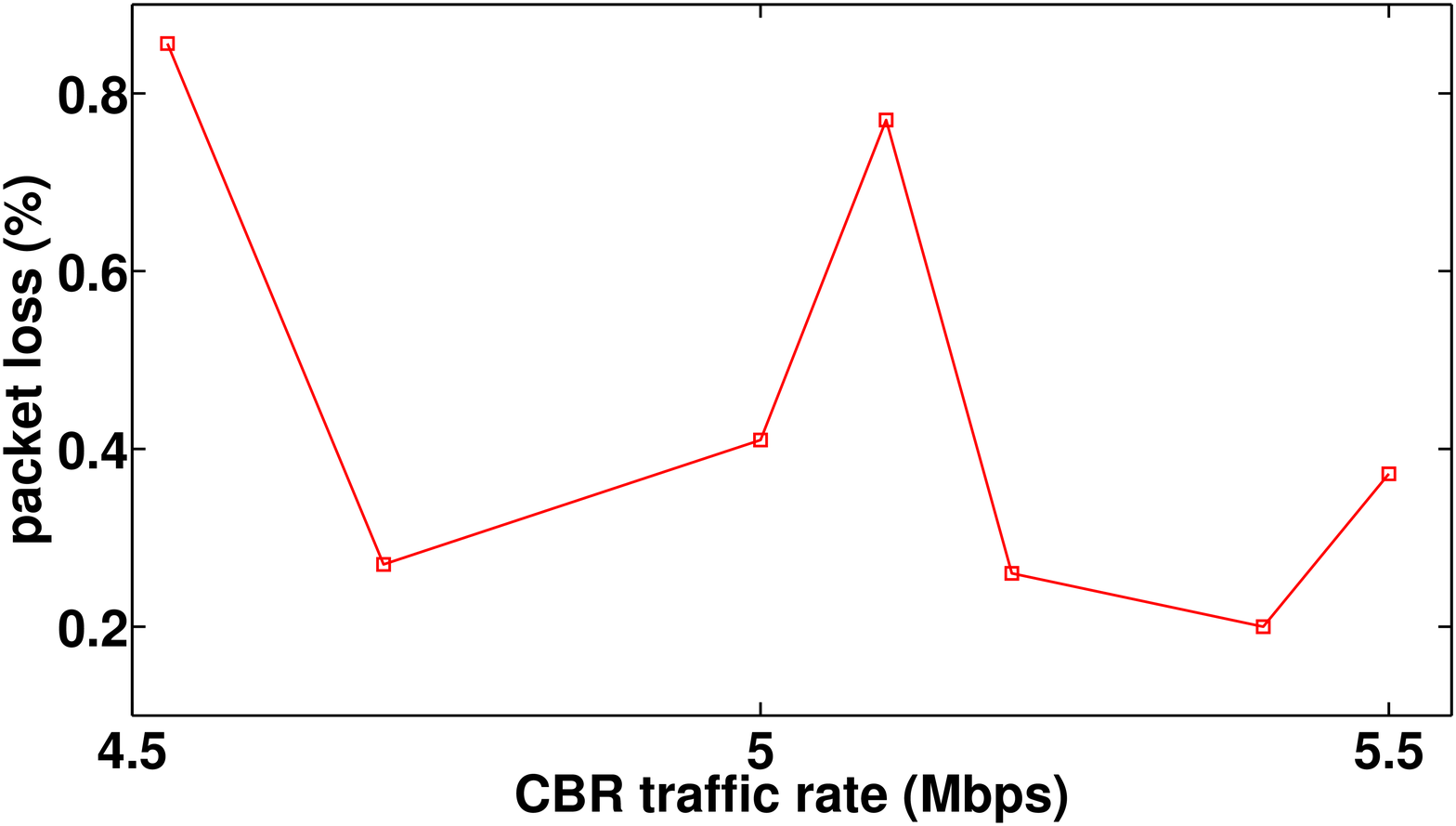}
\caption{Packet losses recorded on the backward channel for a case where the telehaptic packet sizes
are comparable to the TCP packet sizes.}
\label{fig:lossCyber}
\end{minipage}
\end{figure}



\subsubsection{Haptic Jitter}
\label{subsubsec:jitter}
We now turn to measurements of the maximum haptic
jitter. Figure~\ref{fig:jitterPhantom} shows the maximum haptic jitter
curves, both by analysis and simulation, for $R \in [R_b, 5.5$
Mbps$]$. It can be seen that the analytical estimates corroborate well
with the simulation measurements, thereby validating our
characterization. It is to be noted that the indicator variable
$I_{(T_h > \frac{nS_{tcp}}{\mu})}$ in Equation~\eqref{equ:jitter} takes
the value 0 for the chosen setting, i.e., $T_h < nS_{tcp}/\mu$. In
order to test the robustness of our model to the network parameters,
we choose another setting that results in
$I_{(T_h > \frac{nS_{tcp}}{\mu})} =$ 1. Specifically, we use $\mu$ as the
control parameter and vary it in the range [9, 25] Mbps. Also, we set
$R_{cross} =$ 6.9 Mbps, so that $R =$ 8Mbps. In
Table~\ref{table:jittertable}, we report the maximum haptic jitter by
analysis (A) and simulation (S). Throughout the considered range of $\mu$,
it can be seen that our analysis accurately estimates the
maximum haptic jitter.

\ignore{
  It can be noticed that the maximum haptic jitter remains comfortably
  within the QoS limit of 10 ms for the settings chosen so far,
  wherein the TCP source parameters remained unchanged.  In order to
  test the robustness of our model, we now vary the TCP
  parameters. Consider the network setting corresponding to
  Figure~\ref{fig:jitterPhantom}. We set $n =$ 5 and $S_{tcp} =$ 1.5
  kB.  For this case, the maximum haptic jitter turns out to be 12 ms,
  which clearly violates the corresponding QoS specification of 10 ms.
  This suggests that it is crucial to ensure that the cross-traffic
  parameters are configured in a manner that is favorable for haptic
  jitter in order to maintain QoS compliance.}



\subsubsection{Packet Loss} 
\label{subsubsec:loss}

We now report the the packet loss suffered by the telehaptic
stream. Interestingly, for all our simulations reported so far, we
notice that telehaptic packet losses are \emph{zero} in spite of the
regular queue overflows induced by TCP. The rationale behind this
interesting behavior is that the telehaptic source generates smaller
packets compared to the TCP packets (137 bytes per packet on the
backward channel for the telehaptic stream, versus 578 bytes per
packet for the TCP stream). As a result, even when the queue drops a
TCP packet, the adjacent telehaptic packets can still (potentially) be
accommodated in the queue. This observation is in line with the
results in \cite{ref:sawashima}, which also investigates CBR loss in the
presence of TCP cross-traffic.

To confirm our conjecture that smaller telehaptic packet sizes are
responsible for the absence of telehaptic packet losses, we simulate a
scenario with higher resolution haptic, audio, and video devices, so
that the telehaptic packet size becomes comparable to the TCP packet
size. Specifically, consider a haptic device like Cybergrasp
\cite{ref:cybergrasp} or Festo's exohand \cite{ref:festo}.
Assuming two interaction points for each of the ten fingers
of the hands results in a twenty-fold increase in the haptic payload
rate. Additionally, we simulate high-definition audio and video
payload with rates of 128 kbps and 2 Mbps, respectively. This results
in $R_b =$ 4.528 Mbps, and a packet size of 566 bytes on the backward
channel for every millisecond. Note that the telehaptic packets are now
comparable in size to the TCP packets. Figure~\ref{fig:lossCyber} presents the
packet loss (in \%) encountered by this telehaptic stream, where we
vary $R_{cross}$ to get $R$ in the range $[R_b,$ 5.5 Mbps]. Note that
with the larger telehaptic packets, losses do occur. While the
measured telehaptic losses pose no threat to haptic media (with a QoS
limit of 10\%), audio and video (which have a more stringent limit of
1\%) are more susceptible to QoS violations.


To summarize, if telehaptic packets are small relative to TCP packets,
the telehaptic stream sees little or no packet loss. However, if the
telehaptic packets become comparable in size to TCP packets (due to
higher fidelity media devices), packet losses become noticeable.

In conclusion, we see that for QoS-compliant communication under
CBR-based telehaptic protocols, the following conditions need to be
satisfied.
\begin{itemize}
\item For buffer stability, we naturally require that the aggregate
  CBR data rate is less than the link capacity, i.e., $R < \mu.$
\item In order to satisfy the haptic delay constraint, we need
  $d_{max} < 30$ ms, i.e., 
  \begin{equation}
    \label{eq:delay_compliance}
    \tau + \frac{B}{\mu} < 30. 
  \end{equation}
\item In order to satisfy the haptic jitter constraint, we need to
  configure the source and the network parameters such that $\delta_{h(max)} <$
  10 ms, where $\delta_{h(max)}$ is given by Equation \eqref{equ:jitter}.
\item To avoid loss in the presence of concurrent TCP traffic, the
  packet sizes used by the telehaptic protocol should be small
  relative to the TCP packet sizes.
\end{itemize}

\subsection{Network Experiments}
\label{subsec:expts}
In order to validate our model under real network conditions and
with real implementations of TCP NewReno in Debian operating system
(kernel version 4.14), we perform experiments on a
real network setup using the single bottleneck network topology that
we used earlier (shown in Figure~\ref{fig:topology}). Each node in the
topology is run on a virtual machine created using the VMware
virtualization infrastructure. We install the
virtualization software on two workstations, each of which hosts
multiple virtual machines.  The virtual machines corresponding to the
traffic sources and the adjoining intermediate node ($n_2$) are
installed on the same workstation, and the remaining nodes are
installed on another workstation. The two workstations are distantly located
from each others, and are connected via a physical network.

The network parameters are configured as per our simulation settings
($\mu =$ 6 Mbps, $\tau =$ 8 ms, and $B =$ 14 kB) using the NetEm tool
which is a standard built-in Linux kernel feature for network emulation.
The CBR traffic (telehaptic
and cross-traffic) is generated using socket programs running on
the corresponding sources, whereas the TCP NewReno traffic is
generated using the Iperf tool \cite{ref:iperf}. We note that the TCP
NewReno implementation uses minimum packet size of 1512 bytes (further
details explained in Section~\ref{subsubsec:jitterExpts}), and $n =$ 2.


\subsubsection{Haptic Delay}
\label{subsubsec:delayExpts}
We begin by presenting the minimum and the maximum haptic delay
measurements in our network experiments.  As in the measurements
corresponding to Table~\ref{table:hapDelay}, we vary $R_{cross}$ to
get $R$ in the range [$R_b, 5.5$ Mbps]. In
Table~\ref{table:hapDelayExpts}, we report $d_{min}$ and $d_{max}$
measured in our network experiments (E). For ease of comparison, we also
present the delays corresponding to analysis (A) and simulations
(S) that we reported earlier in Table~\ref{table:hapDelay}.
It can be seen that the delay measurements corroborate well with
the analytical estimates, thereby validating the accuracy of our delay
analysis model under real network conditions as well. Note that the
Linux implementation of TCP NewReno uses larger packet sizes, and hence
the queue starts to drop packets at a lower queue occupancy than in
simulations. This results in a lower $d_{max}$ in network experiments.
\begin{table}[h]
\parbox{.55	\linewidth}{
\centering
\resizebox{7.8cm}{1.2cm}{
\begin{tabular}{|c|c|c|c|c|c|c|}%
      \hline
      \multirow{2}{*}{$R$ (Mbps)} & \multicolumn{3}{c|}{$d_{min}$ (ms)} & \multicolumn{3}{c|}{$d_{max}$ (ms)}\\
      \cline{2-7}
        & A & S & E & A & S & E\\ \hline
       1.096 & 9.91 & 8.89 & 8.34 & 26.66 & 26.47 & 22.17 \\ \hline
       2 & 10.74 & 9.21 & 8.79 & 26.66 & 26.40 & 22.58 \\ \hline
       3 & 12.27 & 11.71 & 11.46 & 26.66 & 26.62 & 23.29 \\ \hline
       4 & 14.81 & 12.95 & 12.42 & 26.66 & 26.45 & 23.09 \\ \hline
       5 & 19.87 & 16.95 & 17.13 & 26.66 & 26.55 & 24.58 \\ \hline
       5.5 & 22.68 & 19.77 & 20.06 & 26.66 & 26.43 & 24.69 \\ \hline
\end{tabular}}
    \captionof{table}{Comparison of $d_{min}$ and $d_{max}$ by analysis (A), simulation (S),
    and real network experiments (E) for a wide range of $R$.}
    \label{table:hapDelayExpts}
  }
  \hfill
  \parbox{.4\linewidth}{

  \centering
  \resizebox{3.7cm}{1.2cm}{
  \begin{tabular}{|c|c|c|c|}
  \hline
  \multirow{2}{*}{$R$ (Mbps)}  & \multicolumn{2}{c|}{$\delta_{h(max)}$ (ms)} \\ \cline{2-3}
   & A & E \\ \hline
    1.096 & 5.43 & 10.62\\\hline
    2 & 5.63 & 10.94\\\hline
    3 & 5.83 & 11.27\\\hline
    4 & 6.03 & 11.44\\\hline
    5 & 6.23 & 7.87\\\hline
    5.5 & 6.23 & 6.03 \\\hline
      \end{tabular}}
      \caption{Analytical (A) and experimental (E) measurements for $\delta_{h(max)}$
      over a wide range of $R$.}
      \label{table:jittertableExpts}
      }
\end{table}

\subsubsection{Haptic Jitter}
\label{subsubsec:jitterExpts}
We now move to maximum haptic jitter measurements. As shown in
Equation~\eqref{equ:jitter}, $\delta_{h(max)}$ has a direct dependence
on $S_{tcp}$. We compute the $\delta_{h(max)}$ with $S_{tcp} =$ 1512 B
for different values of $R_{cross}$ using Equation~\eqref{equ:jitter}.
In Table~\ref{table:jittertableExpts}, we present the analytical estimates (A)
and their corresponding experimental measurements (E) of $\delta_{h(max)}$.
As can be seen, there exists considerable error between the two
quantities.

We now explain the rationale behind this error.  From our traces, we
notice that the TCP NewReno implementation in Debian operating system
deviates from the definition in the RFC \cite{ref:newreno} as follows:\\
$S_{tcp}$ is implemented as a dynamic parameter that depends on the
value of $W_{min}$. When $W_{min}$ is low, which corresponds to low
bandwidth availability in the network, the source transmits packets of
(smaller) size $S_{tcp}=$ 1512 B. On the other hand, when $W_{min}$ is
high, which corresponds to high bandwidth availability, the source
increases the packet size to $S_{tcp}=$ 2960 B. For intermediate
values of $W_{min},$ the source transmits a mix of small and large
packets.

As a consequence, the amount of traffic (in bytes) injected by the TCP
source into the network at a slot boundary, which as we have seen
determines $\delta_{h(max)}$, depends on $W_{min}$. We can now
explain the deviation of $\delta_h (E)$ from $\delta_h (A)$ as
follows.
When $R$ is small, we expect $W_{min}$ to be high (as per
Equation~\eqref{equ:wmin}). Therefore, at the slot boundaries the
source transmits three larger packets resulting in higher $\delta_{h(max)}$. It
is worth noting that for $R \in$ [1.096, 4] Mbps, simply by assigning
$S_{tcp}=$ 2960 B, $\delta_{h(max)}(A)$ precisely matches $\delta_{h(max)}(E)$. As $R$
is increased further, the source gradually starts to transmit smaller
packets. For example, when $R =$ 5 Mbps, we observe that one larger
and two smaller packets are transmitted at the slot boundaries. This
results in diminishing error between
$\delta_{h(max)} (E) \text{ and } \delta_{h(max)} (A)$. When $R =$ 5.5~Mbps, three
smaller packets are transmitted, resulting in a negligible error between the two.

We conclude that the maximum haptic jitter is highly sensitive to the
implementation aspects of TCP NewReno. While our analysis assumes the dynamics
specified in the RFC \cite{ref:newreno}, the actual implementation in the Debian
operating system deviates from the RFC, resulting in a mismatch between
the analytical and the experimental jitter. Coming up with an analytical bound
for the haptic jitter that is robust to the specifics of common TCP
implementations is an interesting avenue for future work.


\subsubsection{Packet Loss}
\label{subsubsec:lossExpts}
The telehaptic packet losses in all of our network experiments are
zero. Note that this is because the bottleneck queue can still admit
smaller sized telehaptic packets while it drops the larger sized TCP
packets. Note that the TCP packets here are larger compared to those
in simulations. 

\section{Adaptive Sampling based Telehaptic Protocols}
\label{sec:tcpweber}

In this section, we study the interplay between adaptive sampling
based telehaptic protocols and heterogeneous cross-traffic involving
TCP and CBR flows. An adaptive sampling scheme based protocol
transmits only \textit{perceptually significant} haptic samples on the
forward and/or backward channels.
As before, the goal of this
section is to evaluate the impact of network cross-traffic on telehaptic traffic generated
by the adaptive sampling based telehaptic protocols. This also leads to
formulating the conditions for QoS-compliant telehaptic communication.

If the protocol employs Weber sampler \cite{ref:perception}, a
specific type of adaptive sampling strategy, on the backward channel, it
must also specify how the irregularly spaced, perceptually significant haptic samples are
multiplexed with audio/video data. For a working example, we consider
the visual-haptic multiplexing protocol \cite{ref:vhmux}, which multiplexes
haptic and video streams on the backward channel as follows: The perceptually significant haptic
samples are packetized with video data of worth 1 ms, so that the haptic
samples suffer minimal packetization delay. On the other
hand, when a series of haptic samples are perceptually insignificant,
the protocol packs a large chunk of a video frame, not exceeding data of
worth 15 ms, into a single packet for transmission.


In order to evaluate the protocol with realistic data, we record ten
pilot signals collected from Phantom Omni device during a real
telehaptic activity. For brevity, we report the results only for one of
these traces, but we note that our findings remain consistent across traces.
The video payload rate is set to 400~kbps, as before. We use
the network settings described previously in Section~\ref{sec:validation}.

In Figure~\ref{fig:vhmuxrate}, we plot the instantaneous telehaptic
transmission rate on the backward channel due to visual-haptic multiplexing protocol. As can
be seen from the figure, the instantaneous rate exhibits large fluctuations in
the range [613, 1079]~kbps, while the long term average rate of 712~kbps
is substantially lower compared to the peak instantaneous rate (1079~kbps).
We now move to our investigation of the interplay between this telehaptic flow and network
cross-traffic. We begin by investigating the impact of CBR cross-traffic
alone on the telehaptic traffic (Section~\ref{subsec:vhmuxcbr}), and then
move to heterogeneous cross-traffic case (Section~\ref{subsec:vhmuxcbrtcp}).

\subsection{CBR Cross-Traffic}
\label{subsec:vhmuxcbr}

\begin{figure}[t]
\begin{minipage}[b]{0.5\linewidth}
\centering
\includegraphics[height=3.8 cm,width = 7.5 cm]{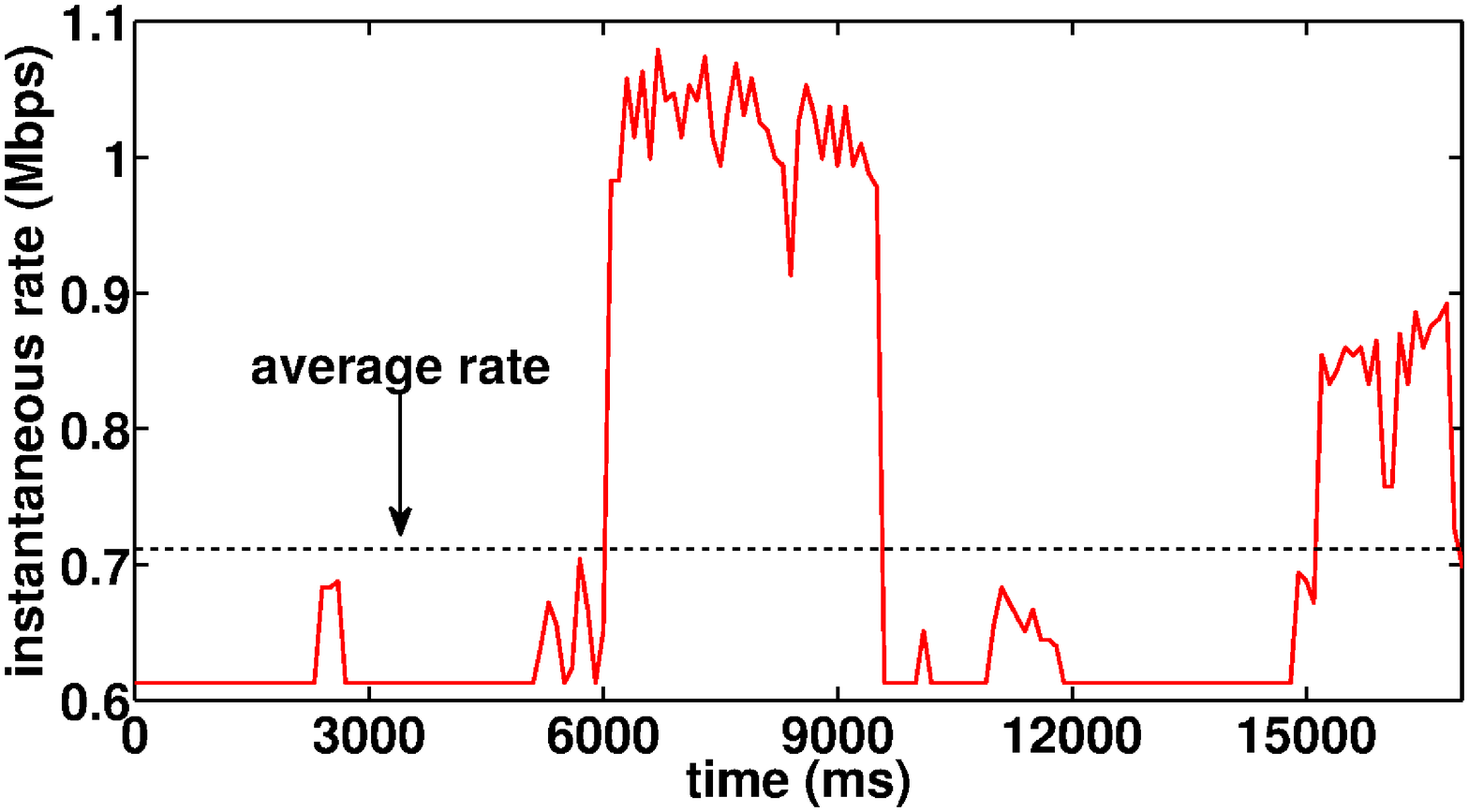}
\caption{Instantaneous telehaptic data rate exhibiting rapid fluctuations under visual-haptic multiplexing.}
\label{fig:vhmuxrate}
\end{minipage}
 \hspace{0.05cm}
\begin{minipage}[b]{0.5\linewidth}
\centering
\includegraphics[height=3.8 cm,width = 7.5 cm]{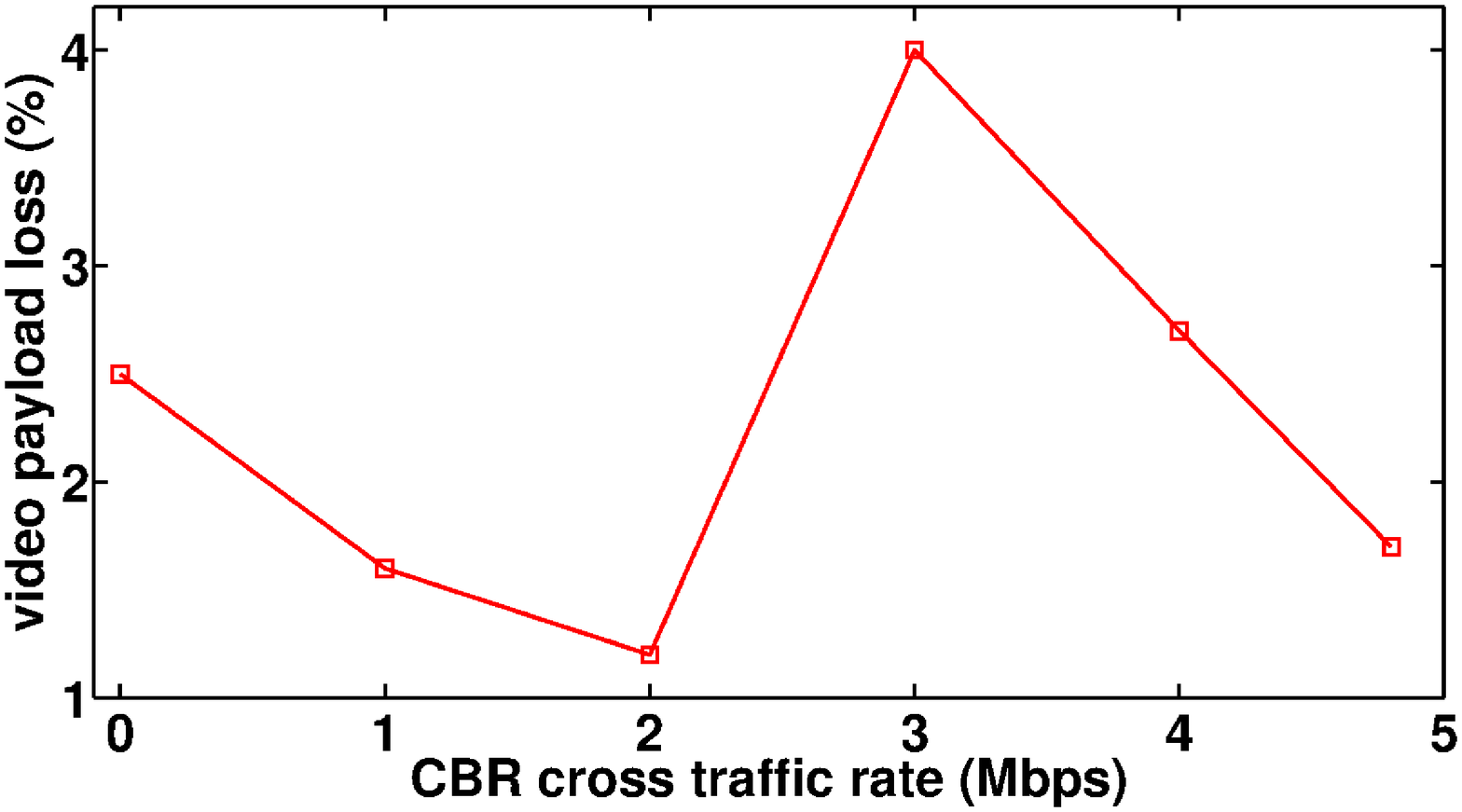} 
\caption{Video payload loss in presence of heterogeneous cross-traffic under visual-haptic multiplexing.}
\label{fig:lossVHMux}
\end{minipage}
\end{figure}

In this section, our goal is to demonstrate that from the standpoint of QoS
compliance on a shared network, the statistical compression provided
by adaptive sampling offers no meaningful economies in terms of
network bandwidth requirement of the telehaptic application.
In other words, the network has
to be able to support the \emph{peak} transmission rate of the
telehaptic flow for QoS compliance.

To illustrate this, we consider an example where the network is provisioned
for the long term average telehaptic rate. This means that the amount of
network bandwidth available for the telehaptic stream at all times
exceeds its average rate. To simulate this
scenario, we set $R_{cross} =$ 5.28 Mbps so that the bandwidth
available to the telehaptic stream is 720 kbps which is greater than the
average rate of 712 kbps, as shown in Figure~\ref{fig:vhmuxrate}.
We remove the TCP source for this experiment. In
Figure~\ref{fig:vhmuxrate}, consider the interval between 6000 ms
and 10000 ms, when the instantaneous rate exceeds the available
bandwidth.
In this interval, our simulation traces reveal a significant haptic
and video payload losses of around 6.2\%. Even though the haptic loss
is below the QoS limits (10\%), the video loss is alarmingly high,
causing severe violations of the QoS requirement (1\%). Additionally,
we note that the haptic delay severely violates the Qos limit of 30 ms.
As another example, setting $\mu =$ 3 Mbps and $R_{cross} =$ 2.28 Mbps
results in larger haptic
and video payload losses of around 9.6\%.

This suggests that for Qos compliance, the network needs to be
provisioned for the peak telehaptic rate rather than the long term
average rate. Hence, the statistical (but network-unaware) compression
achieved by adaptive sampling is not particularly effective from the
standpoint of reducing the bandwidth requirement of the telehaptic
application. This departure from the existing theories on adaptive
sampling schemes, for example \cite{ref:perception,ref:telepresence},
is another key contribution of this work, since the previous studies
treat the long term average data rate as the primary parameter
in the evaluation of quality of a telehaptic interaction.



\subsection{Heterogeneous Cross-Traffic}
\label{subsec:vhmuxcbrtcp}
For the case of heterogeneous cross-traffic, we reinstate the TCP
source on the backward channel. Since $R_b \in$ [613, 1079] kbps, we
vary $R_{cross}$ in the range [0, 4.4] Mbps so that $R \in$ [$R_b$, 5.5 Mbps].
Due to space limitations, we only state our main findings:
\begin{enumerate}
\item Equation~\eqref{eq:delay_compliance} captures the peak delay
  seen by telehaptic packets accurately.
\item Figure~\ref{fig:lossVHMux} shows the video payload loss (in \%)
  recorded for various values of $R_{cross}$. It can be seen that the video loss
  faces severe QoS violations throughout.
  This is because the visual-haptic multiplexing protocol
  transmits large packets with video payload in the absence of
  perceptually significant haptic samples. Recall, from our discussion
  in Section~\ref{subsubsec:loss}, that large packets are more likely
  to get dropped in the presence of TCP cross-traffic. Interestingly,
  haptic media suffers \emph{zero} losses in this case. This is
  because the protocol transmits the perceptually significant samples
  in smaller packets of size 137 bytes. Once again, this illustrates
  that the packet sizing performed by a telehaptic protocol plays a
  crucial role in influencing the loss experienced in the presence of TCP
  cross-traffic.
\end{enumerate}

Note that we do not discuss jitter in this section, since haptic jitter is
harder to define under adaptive sampling, which only transmits
perceptually significant samples that are irregularly placed in time.

To summarize, the conditions for QoS compliance for adaptive sampling
based telehaptic protocols are:
\begin{itemize}
\item The network be provisioned for the peak telehaptic rate in order
  to alleviate the effects of the large fluctuations in the
  instantaneous telehaptic rate.
\item To meet the haptic delay constraint,
  Equation~\eqref{eq:delay_compliance} be satisfied.
\item To avoid packet loss in presence of a TCP flow, large packet
  sizes be avoided.
\end{itemize}

%
%
%

\section{Delay QoS Compliance for Audio-Video}
\label{sec:compliance}

While the previous sections have largely focused on haptic QoS
compliance, we consider audio/video QoS compliance in this
section. The goal of this section is to show that meeting the haptic
delay requirement for CBR based telehaptic protocols typically
guarantees that the (less stringent) delay requirements for audio and
video are also satisfied under reasonable multiplexing schemes.

Since the relationship between haptic delay and audio-video delay
depends strongly on the multiplexing framework being employed, we consider a specific
example -- the hierarchical multiplexing scheme developed in
\cite{ref:dpm}; an analogous analysis can also be performed for other
multiplexing mechanisms. The multiplexing mechanism in \cite{ref:dpm}
is the following: An audio/video fragment of fixed size $s_m$ (in bytes) is transmitted
along with each haptic sample, with audio payload having strict
priority over video payload. With the assumption that the haptic delay deadline of 30 ms
is met, we can derive the following expressions for the maximum delay
experienced by the audio and the video frames.
\begin{equation*}
  d_{aud} = 30 + \frac{s_a}{s_m}T_h, \qquad d_{vid} = 30 + \frac{1}{f_v},
\end{equation*}
where $s_a$ (in bytes) is the size of an audio frame, $f_v$ (in Hz) denotes
the frame rate of the video signal, and $T_h$ is the inter-packet gap of telehaptic
stream as before; see the reference \cite{ref:dpm} for more details on this.

\ignore{
We now move to obtaining expressions for maximum audio and video
delays assuming that the haptic media meets its corresponding delay
deadline.  We note that the delays encountered by the audio and video
frames depend heavily on the multiplexing scheme being used. For the
purpose of illustration, we consider a specific type of hierarchical
multiplexing scheme described in \cite{ref:dpm}, where the audio/video
data is transmitted along with a haptic sample at the rate of 1000
packets/sec.  The multiplexer gives strict priority to audio over
video. Under the condition that the haptic delay QoS of 30 ms is met,
we derive expressions for the maximum delays experienced by the audio
and the video frames. Through simple analysis, it can be shown that
the maximum audio and video delays (in ms) for this multiplexer are
respectively given by
\begin{equation*}
d_{aud} = 30 + \frac{S_a}{S_m}(1ms), \qquad d_{vid} = 30 + \frac{1}{f_v}
\end{equation*}
where $s_a$ is the size (in bytes) of an audio frame, and $s_m$ is the
size (in bytes) of the audio data transmitted in each packet. $f_v$
denotes the frame rate (in Hz) of the video signal.  }

For the setting considered in Section~\ref{sec:validation}, it can be
shown that $s_a =$ 160 bytes, $s_m =$ 58 bytes, and $f_v =$ 25
Hz. Hence, $d_{aud} =$ 32.75 ms and $d_{vid} =$ 70 ms.  We see that
even the worst case audio and video delays are comfortably within
their respective QoS limits.
Thus, we conclude that compliance with the haptic delay
constraint (which follows from Equation~\eqref{eq:delay_compliance}) implies
compliance with the delay constraints for audio and video media as well.

\section{Related Work}
\label{subsec:relatedwork}
In this section, we discuss the prior works relevant to this
paper. Surprisingly, we note that there are no works that make a
detailed examination of the interplay between telehaptic and TCP
flows.
A few works, however, have included TCP flows in their investigation,
but the analyses themselves are rather trivial to draw any broad
conclusions \cite{ref:wirzetp,ref:opportunistic}. In the rest of this
section, we present a brief review of the literature focused on the
interplay between generic UDP (voice or video) and TCP flows.

\subsubsection{Impact on TCP Flows:}
A large volume of work is present in the literature in which the sole
performance metric is TCP throughput. The work in \cite{ref:tcpudp4}
provides an understanding of the network bandwidth sharing between
multimedia streaming and TCP flows. In
\cite{ref:zahedi,ref:DoS,ref:impactDoS,ref:bruno}, the authors discuss
the impact of UDP flows on TCP throughput in wireless adhoc networks
or LANs. The authors in \cite{ref:tcpudp1}
present a novel mechanism to enhance the TCP throughput in presence of
concurrent UDP flows. The work in \cite{ref:rohner} demonstrates that the
interaction of UDP and TCP on multihop wireless networks can result in
significantly low throughput and unstable routes. The authors in
\cite{ref:suznjevic} investigate the effect of UDP flows on the online
gaming flows that are TCP-based, again with an emphasis on TCP
throughput. The authors in \cite{ref:beritelli,ref:tian} investigate
the impact of Voice over IP (VoIP) traffic on TCP traffic. A few works
have proposed protocol designs for UDP-based traffic that yield to TCP
flows in a fair manner; see, for example,
\cite{ref:rejaie,ref:rhee,ref:handley}. It is important to remark that
none of the prior works discussed so far focus on the impact of TCP
traffic on the QoS experienced by the UDP flow.

\subsubsection{Impact on UDP Flows:}
As the multimedia streaming and tele-conferencing applications (both typically use UDP) gradually started gaining popularity, studies concerning the impact of TCP on UDP flows became the center of gravity for many research groups. The primary performance metrics of interest in these studies are the following: delay, jitter, and packet loss. In the rest of this section, we systematically discuss the prior works considering each of these metrics one-by-one.\\

\vspace{-2mm}
\noindent\textbf{Delay:}
The work in \cite{ref:veres} explores the possibility of providing
differentiated services to voice, video, and data traffic with an
objective of guaranteeing simultaneous delay QoS-compliance to media
applications in a wireless network setting. A non-exhaustive list of
works that carry out further investigations in this direction are
\cite{ref:arranz,ref:shetiya,ref:zhai,ref:papadimitriou,ref:boggia,ref:andreadis,ref:kim}. In
\cite{ref:xiang}, the authors employ a Markov chain model to estimate
the average delay when UDP traffic shares the network resources with
TCP streams on a wireless LAN. Recent investigations in this direction
\cite{ref:decicco,ref:zhang} employ the popular tele-conferencing tool
Skype (which uses UDP) for studying its interplay with TCP
traffic. These works report the long-term average
RTT encountered by the Skype packets. Note that in each of the above
studies, only the average delay is considered as the performance
metric for real-time voice and/or video traffic. However, we remark
that for ultra-sensitive telehaptic applications the instantaneous
delay is a more meaningful evaluation metric than the time-average delay.

A few recent studies report the instantaneous delay of UDP packets
under the influence of TCP flows. The authors in \cite{ref:xuvideo}
study voice and video delays with popular video telephony applications
Google+, iChat, and Skype. Similar investigations have been carried
out with a recently proposed protocol named Google Congestion Control (GCC) in
\cite{ref:gcc,ref:carlucci}. It is worth noting that all of the above works
merely report the measured instantaneous UDP delays for the considered
network settings.

To summarize, all of the above studies take into consideration only the experimental measurement of the UDP delays. It is imperative to point out that a theoretical characterization that provides a general model for the observed UDP delay profiles is lacking in the literature. Further, hitherto no work has considered investigating the conditions for meeting the delay deadline for haptic modality, which is more challenging compared to audio and video delay criteria.\\

\vspace{-2mm}
\noindent\textbf{Jitter:}
Several works like \cite{ref:bonald,ref:jitter1,ref:jitter2} have attempted to examine the time-average jitter suffered by a generic UDP streams due to coexisting TCP cross-traffic. Under similar settings, the authors in \cite{ref:jitter3} have reported the average jitter encountered by voice and video streams. However, similar to delay, the more relevant parameter of interest is the instantaneous jitter which the above works do not explore.
The reference \cite{ref:jitter4} presents the instantaneous jitter faced by video streams in presence of TCP and other UDP streams. However, their investigation is based only on experimental observations; a mathematical model for characterization of the jitter is missing.

The authors in \cite{ref:jitter5} experimentally investigate the impact of different packet scheduling schemes, like priority queueing and weighted round robin, on video jitter. In contrast, in this article, we consider droptail scheduling, which is a widespread form of packet scheduling in the internet. The work in \cite{ref:jitter7} develops a statistical model for simulating jitter behavior in packet networks. However, both these works do not take the TCP rate dynamics into account.\\

\vspace{-2mm}
\noindent\textbf{Packet Loss:}
The first known effort in analyzing the UDP packet losses under the
effect of TCP streams was carried out in \cite{ref:sawashima}. Their
hypothesis is that when the network queues are full, smaller UDP
packets have a lower likelihood of getting dropped, and
vice-versa. The work in \cite{ref:xylomenos} carries out similar
investigation.

Several other works also analyze the UDP packet losses using a variety
of control parameters. While the works in
\cite{ref:bonald,ref:papadimitriou} use the number of competing
streams (network load), the works in
\cite{ref:vishwanath,ref:bai,ref:zhang} study UDP losses from the
standpoint of the queue sizes. The authors in \cite{ref:hasslinger}
infer the relationship between packet transmission timescales and UDP
losses.

More recent studies have investigated the losses induced by TCP on
VoIP flows. The work in \cite{ref:decicco} studies Skype packet losses
under time-varying network bandwidth conditions. Similar experiments
have been conducted with GCC \cite{ref:gcc}.  We note that these works
only report the observed packet loss without shedding light on its
dependence on any parameter.

Since we seek to investigate the vulnerability of the telehaptic
packets to queue drops, the work most closely related to our work is
\cite{ref:sawashima}.

\section{Concluding Remarks}
\label{sec:conclusions}

In this paper, we presented a comprehensive assessment of the
interplay between telehaptic protocols and heterogeneous cross-traffic
in a shared network. For CBR based telehaptic protocols, we derived
bounds on the delay as well as jitter, whose accuracy was validated
through extensive simulations as well as network experiments. For
adaptive sampling based protocols,
we observed that the network should be provisioned for the peak
telehaptic rate to prevent QoS violations. Our analysis and
experiments lead us to formulate a set of conditions for QoS-compliant
telehaptic communication on shared networks. These conditions can in
turn be used to characterize the class of network settings where QoS
compliant telehaptic communication is feasible.

\ignore{In this paper, we presented a comprehensive mathematical
  model of the interplay between CBR and TCP traffic streams that are
  commonly served by the shared networks. This generic model can be
  directly applied to any scenario where a CBR flow shares network
  resources with a TCP flow. Further, we applied this model to the
  case of telehaptic traffic, produced by two classes of protocols,
  and heterogeneous traffic co-existing on the network in order to
  gain insights on the impact of cross-traffic on telehaptic QoS. This
  led to the formulation of a set of sufficiency conditions for QoS
  compliance of haptic delay and jitter.  We conducted extensive
  simulations and network experiments for the validation of our
  characterizations.  Additionally, we discovered two important,
  experiment-driven conditions for QoS compliance. First, in order to
  ensure a near-zero packet loss the telehaptic packets need to be
  small relative to the TCP packet size. Second, when employing the
  adaptive sampling for telehaptic communication the network should be
  provisioned for the peak telehaptic rate to prevent QoS
  violations. Finally, we showed that meeting delay QoS needs for
  haptic media is a sufficient condition for guaranteeing audio and
  video delay QoS compliance.}

\bibliographystyle{ACM-Reference-Format-Journals}
\bibliography{cbrtcp}

\newpage
\section*{ONLINE APPENDIX}
\label{sec:appendix}

\section{Analysis of TCP-CBR Interplay}
\label{append:analysis}

In this section, we report in detail the development of the analytical
model that characterizes the dynamics of interplay between TCP and CBR
traffic on a shared network. We begin by characterizing the queue
occupancy (Section~\ref{append:queue}), and then move to
characterization of the maximum CBR jitter
(Section~\ref{append:jitter}). 

\subsection{Characterization of Queue Occupancy}
\label{append:queue}
We present Figure~\ref{fig:cwndqueue} with a few additional notations
in Figure~\ref{fig:tcpcwndbuffer}. Since we are considering only TCP
and CBR traffic types, the queue occupancy at the onset of
$i$\textsuperscript{th} slot $Q(i) = Q_C(i)+Q_T(i),$ where $Q_C(i)$
and $Q_T(i)$ denote the amount of CBR and TCP traffic in the queue, respectively,
at the onset of $i$\textsuperscript{th} slot.  Let $T$ denote the
duration of the fast retransmit, fast recovery phase.
\begin{figure}[!h]
\centering
\includegraphics[height = 50mm, width = 100mm]{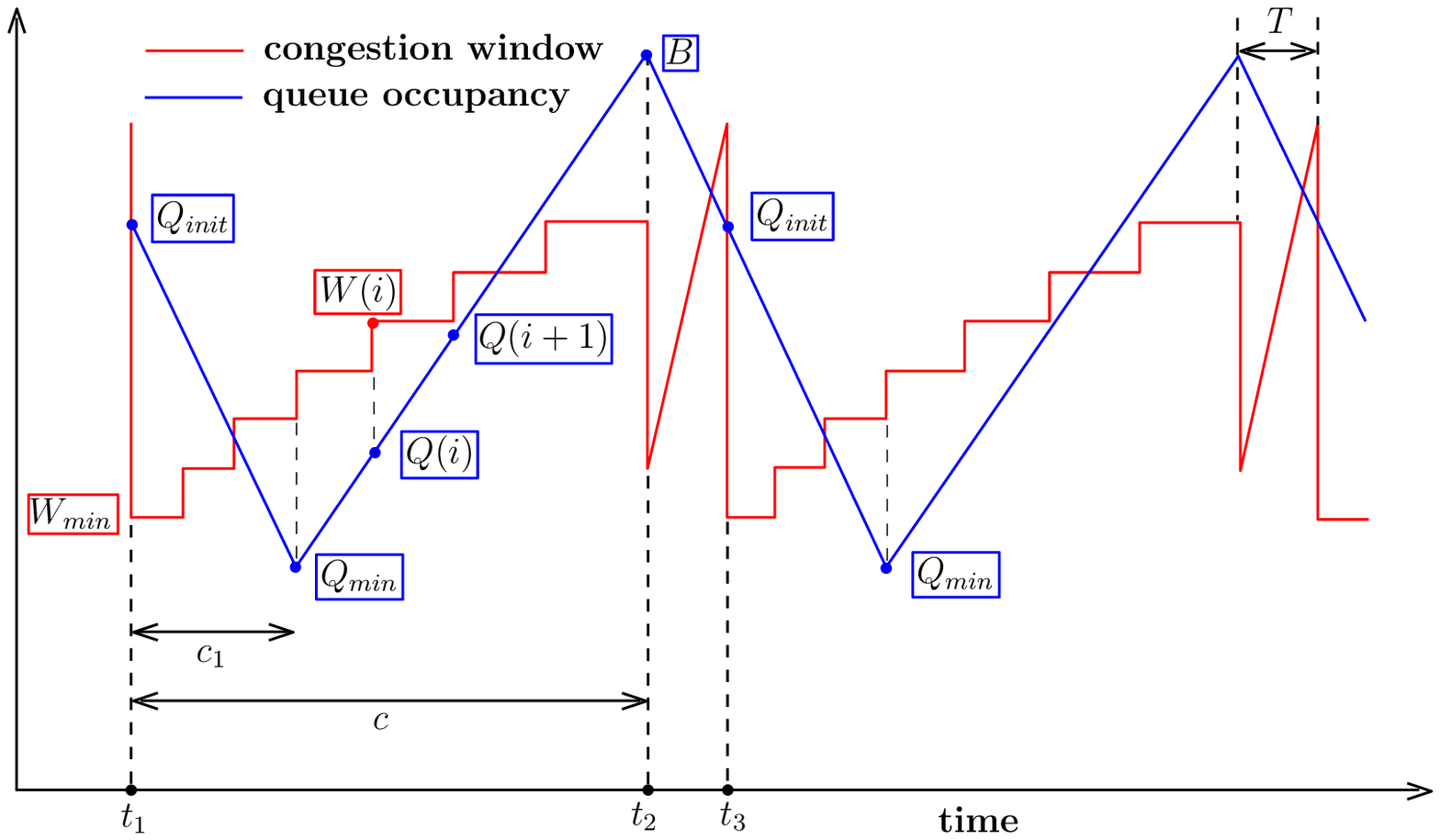}
\caption{Evolution of TCP congestion window and bottleneck queue occupancy for heterogeneous traffic flows involving TCP and CBR streams.}
\label{fig:tcpcwndbuffer}
\end{figure}

We begin by analyzing the queue occupancy evolution in the
congestion avoidance phase (Section~\ref{append:congavoid}), and
subsequently move to the fast retransmit, fast
recovery phase (Section~\ref{append:frfr}).

\subsubsection{Congestion Avoidance}
\label{append:congavoid}
In this section, we analyze in detail the queue dynamics in the
congestion avoidance phase ($t_1$ to $t_2$ in
Figure~\ref{fig:tcpcwndbuffer}).  Unlike the congestion avoidance in
the single TCP source case (depicted in Figure~\ref{fig:cwnd}), where the
queue occupancy varies monotonically, this phase in the TCP-CBR case
can be split into the following two regions:
\begin{enumerate}
 \item \emph{increasing region}: queue occupancy builds up from $Q_{min} \text{ to } B$ over $c-c_1$ slots, 
 \item \emph{decreasing region}: queue occupancy reduces from $Q_{init} \text{ to } Q_{min}$ over $c_1$ slots.
\end{enumerate}
We seek to obtain the relationships between queue occupancy at
various stages in each of the above mentioned regions. \\
\vspace{-2mm}

\noindent\textbf{Increasing region:}
Let $RTT(i)$ denote the duration of the
$i$\textsuperscript{th} slot. Recall that $RTT(i)$ is the RTT encountered by the
probing packet transmitted in the $i$\textsuperscript{th} slot. Therefore, we can write
\begin{equation}
RTT(i) = 2\tau+\frac{Q(i)}{\mu},
\label{equ:rtt}
\end{equation}
where $2\tau$ is the round trip propagation delay, and the second term
is the queueing delay faced by the probing packet at the ingress of the
bottleneck link. Note that $Q(i)$ is the queue occupancy
at the onset of $i$\textsuperscript{th} slot, which is exactly the
queue occupancy seen by the probing packet.

We know that the TCP source injects an amount of traffic equal to
$W(i)S_{tcp}$ over $i$\textsuperscript{th} slot. Let $D_T(i)$ denote
the amount of TCP traffic drained from the queue during
$i$\textsuperscript{th} slot.
Recall that the TCP source transmits an additional packet in each
slot relative to the previous slot.
Based on the analysis in \cite{ref:tcpbuffer}, we make a reasonable assumption
that the TCP component of
queue occupancy builds up at the rate of 1 packet per slot.
Relating initial states of the
queue, input, and output during $i$\textsuperscript{th} and
$(i+1)$\textsuperscript{th} slots, we obtain the following equation for
the TCP component of the queue.
\begin{equation}
\label{equ:tcpinputoutput}
[Q_T(i+1) + W(i+1)S_{tcp} - D_T(i+1)] - [Q_T(i) + W(i)S_{tcp} - D_T(i)] = S_{tcp}, \forall i \in  [c_1+1,c-1].
\end{equation}
Here, the first and the second terms in LHS signify the queue occupancy
at end of $(i+1)$\textsuperscript{th} and $i$\textsuperscript{th} slots, respectively.

We now derive an analogous equation for CBR component of the queue. The
amount of traffic injected by the CBR stream during the $i$\textsuperscript{th} slot is given by
$RTT(i)R$. Note that in the increasing region $RTT(i+1)>RTT(i)$, and hence
the CBR source transmits higher amount of
traffic in the $(i+i)$\textsuperscript{th} slot relative to the $i$\textsuperscript{th}
slot.
Let $\Delta Q_C(i)$ denote the difference in the CBR component of the queue at the end
of $(i+1)$\textsuperscript{th} and $i$\textsuperscript{th} slots.
Let $D_C(i)$ denote the amount of CBR traffic
drained from the queue during $i$\textsuperscript{th} slot.
Relating the initial queue states, input and output during $i$\textsuperscript{th} and
$(i+1)$\textsuperscript{th} slots, we obtain the following equation for the CBR component of the queue.
\begin{equation}
\label{equ:teleinputoutput}
\begin{split}
[Q_C(i+1) + RTT(i+1)R-D_C(i+1)] - [Q_C(i) + RTT(i)R-D_C(i)] = \Delta Q_C(i), \\ \forall i \in  [c_1+1,c-1].  
\end{split}
\end{equation}

\noindent We know that $Q_T(i)+Q_C(i)=Q(i)$, and the total queue drain over the $i$\textsuperscript{th} slot
$D_T(i)~+~D_C(i)~=~\mu RTT(i)$. Using these relationships in Equations
(\ref{equ:tcpinputoutput}) and (\ref{equ:teleinputoutput}), and subsequently adding them
up, we obtain
\begin{equation}
\label{equ:tcpteleinputoutput1}
\begin{split}
Q(i+1)-Q(i) + (R-\mu)[RTT(i+1)-RTT(i)] + [W(i+1)-W(i)]S_{tcp} \\ = S_{tcp}+\Delta Q_C(i), 
\forall i \in  [c_1+1,c-1].
\end{split}
\end{equation}

We know that $Q_T(i+1)-Q_T(i) = S_{tcp}, \forall i \in [c_1+1,c-1]$, and
$W(i+1)-W(i) = 1$.
Hence, the queue occupancy increases at the rate of $\Delta Q_C(i)+S_{tcp}$ per slot
i.e., $Q(i+1)-Q(i) = S_{tcp}+\Delta Q_C(i), \forall i \in [c_1+1,c-1]$. Using these
relationships in Equations~\eqref{equ:rtt} and (\ref{equ:tcpteleinputoutput1}),
we obtain
\begin{equation}
\label{equ:tcpteleinputoutput3}
\Delta Q_C(i) = \frac{R S_{tcp}}{\mu-R}, \forall i \in  [c_1+1,c-1].
\end{equation}
This suggests that in the increasing region of congestion avoidance, the CBR traffic
builds up in the queue at a constant rate of $\frac{RS_{tcp}}{\mu-R}$ per slot.
Further, the ratio of increments in CBR and TCP components per slot
equals $\frac{R}{\mu-R}$. Interestingly, this ratio is independent of
TCP parameters.

We now seek to derive the relationship between the end-to-end queue occupancy
in the increasing region (i.e., $Q_{min}$ and $B$). From Equation~\eqref{equ:tcpteleinputoutput3}
we can calculate that $Q(i)$ increases at the rate of $\frac{\mu S_{tcp}}{\mu-R}$ per slot.
Since the queue occupancy build up from $Q_{min}$ to $B$ comprises of $c-c_1$ slots, we can write
\begin{equation}
\label{equ:queuelength1}
Q_{min}+(c-c_1)\frac{\mu S_{tcp}}{\mu - R} = B.
\end{equation}

\noindent\textbf{Decreasing region:} We now move to the decreasing region of the congestion avoidance, i.e.
$i \in [1,c_1]$.
Note that since the queue occupancy is a decreasing function of $i$ in this region,
we can infer (from Equation~\eqref{equ:rtt}) that RTTs encountered by the
probing packets in successive slots reduce progressively.
Using the basic input-output equation in each of the slots, we can write
\begin{equation}
\label{equ:queuelengthdec}
Q(i+1) = Q(i)+RTT(i)R+W(i)S_{tcp} - \mu RTT(i), \forall i \in [1,c_1].
\end{equation}
By definition, $Q(1) = Q_{init}$, $Q(c_1+1) = Q_{min}$, and $W(i) = W_{min}+i-1$.
Adding up the $c_1$ components of Equation~\eqref{equ:queuelengthdec}, we obtain the
following relationship between $Q_{init}$ and $Q_{min}$.
\begin{equation}
\begin{split}
\label{equ:queuelengthdec1}
Q_{min} = Q_{init} + (R-\mu)\sum\limits_{i=1}^{c_1} RTT(i) + \sum\limits_{i=1}^{c_1} (W_{min}+i-1)S_{tcp}.
\end{split}
\end{equation}
Substituting Equation \eqref{equ:rtt} in \eqref{equ:queuelengthdec1}, we obtain
\begin{equation}
Q_{min} = Q_{init} \alpha^{c_1} + \Bigg(\frac{1-\alpha^{c_1}}{1-\alpha}\Bigg)[W_{min}S_{tcp}-(1-\alpha)2\mu\tau] + S_{tcp}\sum_{j = 0}^{c_1-2} (c_1-1-j)\alpha^j,
\label{equ:queuelengthdec2}
\end{equation}
where $\alpha = R/\mu$. \\



\subsubsection{Fast Retransmit, Fast Recovery}
\label{append:frfr}
We now move to modeling the queue dynamics in the fast retransmit, fast recovery phase (interval between $t_2$ and $t_3$ in Figure~\ref{fig:tcpcwndbuffer}).\\

\noindent\textbf{End-to-end queue occupancy:} It can be shown that over the duration $T$,
the TCP NewReno source transmits $W_{min}$ packets and receives $2W_{min}$ ACKs (including duplicates).
A total of $2W_{min}$ ACK arrivals imply that $2W_{min}$ TCP packets
have escaped the bottleneck link in the duration $T$. Recall (from
Section~\ref{subsec:tcpbg}) that there are $2W_{min}$ outstanding packets at the start of the fast retransmit, fast recovery phase (at $t_2$ in Figure~\ref{fig:tcpcwndbuffer}).
This implies that the fast retransmit, fast recovery phase ends (at $t_3$) exactly when all of
the TCP packets that were outstanding at the beginning (at $t_2$) have been acknowledged. We already know that at $t_2$ the queue
occupancy $B$ is shared between CBR and TCP traffic in the ratio $\frac{R}{\mu-R}$.
This implies that over the duration of $T$, along with $2W_{min}S_{tcp}$ bytes of TCP data, $2W_{min}S_{tcp} \frac{R}{\mu-R}$ bytes of CBR data have also escaped the bottleneck link. Therefore, we obtain the expression for $T$ as
\begin{equation}
\label{equ:T}
T = \frac{2W_{min}S_{tcp} + 2W_{min}S_{tcp}\frac{R}{\mu-R}}{\mu} = \frac{2W_{min}S_{tcp}}{\mu-R}
\end{equation}

\noindent Owing to the transmission of $W_{min}$ TCP packets, we can express the relationship between $B$ and $Q_{init}$
\begin{equation}
\label{equ:qinitT}
Q_{init} = B+(R-\mu)T+W_{min}S_{tcp}
\end{equation}

\noindent From Equations \eqref{equ:T} and \eqref{equ:qinitT}, we obtain
\begin{equation}
\label{equ:qinitfinal}
Q_{init} = B-W_{min}S_{tcp}
\end{equation}

\noindent\textbf{TCP component of queue at the end:} We now seek to compute the amount of TCP
data present in the queue at $t_3$. As discussed previously, all the TCP packets that are outstanding at $t_2$ are acknowledged at $t_3$. This clearance of backlog
implies that all of the TCP packets present in the queue at $t_3$ must have been injected in the interval $T$ (between $t_2$ and $t_3$). At $t_3$, these packets are either present in the queue or are in flight in the channel. For the ease of analysis, we make a reasonable assumption that the ratio of amount
of TCP data present in the queue at $t_3$ and total TCP data injected in $T$ is comparable to the
corresponding ratio of the CBR stream.
Recall that the amount of CBR and TCP data transmitted in the interval $T$ are given as $RT$ and $W_{min}S_{tcp}$, respectively.
Hence, the ratio of CBR and TCP data in the queue at $t_3$ can be written as
$\frac{RT}{W_{min}S_{tcp}}$. Using the expression for $T$ from
Equation (\ref{equ:T}), this ratio reduces to $\frac{2R}{\mu-R}$. 

This finding is striking as the ratio of CBR and TCP components in the queue doubles
during the interval $T$. Recall that this ratio is equal to $\frac{R}{\mu-R}$ at the beginning of this interval. The justification for the increase in the ratio is the following:
As per the TCP NewReno protocol design, after the retransmission of the lost packet, the TCP source makes no transmissions
until it receives $W_{min}$ ACKs. On the contrary, the CBR source continues to pump in data at the steady rate of $R$. Thus, the queue is predominantly occupied by the CBR traffic at the end of $T$, thereby bloating the ratio of CBR and TCP contents in the queue.

Using this ratio, the amount of TCP data in the queue at $t_3$ can be expressed as $Q_{init}(\frac{\mu-R}{\mu+R})$. Naturally,
the channel will be shared between the TCP and the CBR streams in the same proportion as the queue. Hence, the amount of in-flight TCP data can be approximated as
$2\mu\tau(\frac{\mu-R}{\mu+R})$. Therefore, we obtain an equation for the TCP component of $Q_{init}$ as follows.
\begin{equation}
\label{equ:qinitTcpcomp}
Q_{init}\Big(\frac{\mu-R}{\mu+R}\Big) = W_{min}S_{tcp} - 2\mu\tau\Big(\frac{\mu-R}{\mu+R}\Big)
\end{equation}

\subsubsection{Congestion Window}
\label{append:congwind}
We now move to relating the congestion windows at the start and the end of a cycle. As discussed in Section~\ref{subsec:tcpbg}, at
the beginning of a cycle $W$ is set to half its value at the end of congestion avoidance
in the previous cycle. Recall that $W$ is incremented $c-1$ times during congestion
avoidance. Relating the congestion window at the start and the end of congestion avoidance, we obtain
\begin{equation*}
\label{equ:windowupdate}
W_{min} = \frac{W_{min}+c-1}{2}
\end{equation*}
which on simplification gives
\begin{equation}
\label{equ:windowupdate1}
W_{min} = c-1
\end{equation}

\noindent Solving the simultaneous equations (\ref{equ:qinitfinal}),
(\ref{equ:qinitTcpcomp}) and (\ref{equ:windowupdate1}), we obtain the closed form expressions for
$W_{min}$, $Q_{init}$ and $c$ as follows.
\begin{equation}
\label{equ:wminapp}
W_{min} = \frac{(B+2\mu\tau)(1-\alpha)}{2S_{tcp}}
\end{equation}
\begin{equation}
\label{equ:qinitapp}
Q_{init} = \frac{(B+2\mu\tau)(1+\alpha)}{2}-2\mu\tau
\end{equation}
\begin{equation}
\label{equ:capp}
c = \frac{(B+2\mu\tau)(1-\alpha)}{2S_{tcp}} + 1
\end{equation}

\noindent Using these in Equations
\eqref{equ:queuelength1} and \eqref{equ:queuelengthdec1}, we obtain
\begin{equation}
\label{equ:qminfinal1}
Q_{min}+\Big[\frac{(B+2\mu\tau)(1-\alpha)}{2 S_{tcp}}-c_1+1\Big]\Big[\frac{S_{tcp}}{1-\alpha}\Big] = B
\end{equation}
\begin{equation}
\label{equ:qminfinal2}
Q_{min} = \Big[\frac{(B+2\mu\tau)(1+\alpha)}{2}-2\mu\tau\Big]\alpha^{c_1} + \Big[\frac{(B-2\mu\tau)(1-\alpha^{c_1})}{2}\Big] +  S_{tcp} \sum_{j = 0}^{c_1-2} (c_1-1-j)\alpha^j
\end{equation}
This completes the formal derivation of Equation~\eqref{equ:qmin2} that we
briefly described in Section~\ref{subsubsec:queue}.

\subsection{Characterization of CBR Jitter}
\label{append:jitter}
In this section, we present the detailed derivation of analytical
expression for $m_{tcp}$ -- the maximum number of
TCP packets transmitted between two successive CBR packets.

As pointed out in Section~\ref{append:frfr}, during congestion avoidance,
the queue and hence the channel are shared by the CBR and TCP streams
in the ratio $\frac{R}{\mu-R}$ (see Equation~\eqref{equ:tcpteleinputoutput3}).
In other words, the TCP stream gets served at the rate of $\mu-R$. Let us now
consider two adjacent ACKs, none corresponding to probing packets. The time
spacing between the two ACKs can be given as $\frac{nS_{tcp}}{\mu-R}$, since there are
$n$ packet receptions between the transmission of two ACKs. Note that each of these ACKs give rise to transmission of an $n-$packet burst.

On the other hand, the transmission of a probing packet gives rise to a slightly different situation. Recall that the probing packet is always transmitted
as a part of an $(n+1)-$packet burst. When the earliest packet in this burst arrives at the
receiver, two scenarios can occur depending on $p$ - the number of packets waiting at the receiver to be acknowledged. Naturally, $p \leq n-1$. We now consider the different possible values that $p$ can take, and analyze each scenario.
\begin{enumerate}
\item $p < n-1$: In this case, the reception of the earliest $n-p$ packets trigger an ACK.
The remaining $p+1$ packets in the burst wait for the earliest $n-p-1$ packets from the subsequent burst for
triggering the next ACK. In this case, the time interval between these two ACKs can be calculated to be $\frac{nS_{tcp}}{\mu-R}$.
\item $p = n-1$: In this case, the reception of the first packet in the burst triggers an ACK.
The latter $n$ packets of the same burst trigger another ACK. It is worth remarking that since the TCP packet
transmissions are bursty in nature, it is fairly reasonable to assume that the TCP packets
belonging to a burst occupy contiguous locations in the queue. Hence, the TCP packets
consume the full channel capacity ($\mu$) until the burst is served by the queue.
Therefore, the time interval between the two back-to-back ACKs in this case can be calculated as
$\frac{nS_{tcp}}{\mu}$.
\end{enumerate}
The important difference between the above two scenarios is the following: For generating two ACKs, in the latter case a single burst of packets is sufficient, whereas in the former case two bursts need to be necessarily transmitted. Therefore, the ACKs are spaced more closely in the latter case than the former. Hence, the ACKs corresponding to the latter case give rise to maximum number TCP packets injected between two CBR packets. It can be shown that the situation $p=n-1$ is guaranteed to occur once every $n$\textsuperscript{th} window update during congestion avoidance. Hence, for this analysis we restrict our attention to the latter case.

We refer back to Figure~\ref{fig:cumulack} for the current analysis.
If $T_h < \frac{nS_{tcp}}{\mu}$, then $m_{tcp} = n+1$ as
no more than one burst can be transmitted in the interval
$T_h$. On the other hand, if $T_h > \frac{nS_{tcp}}{\mu}$,
the number of $n-$packet bursts transmitted in addition to the
$(n+1)-$packet burst in the interval $T_h$ can be expressed as 
$1+\floor*{\frac{T_h-\frac{nS_{tcp}}{\mu}}{\frac{nS_{tcp}}{\mu-R}}}$.
Hence, we write the general expression for $m_{tcp}$ as
\begin{equation}
m_{tcp} = n+1+\Bigg(1+\floor*{\frac{T_h-\frac{nS_{tcp}}{\mu}}{\frac{nS_{tcp}}{\mu-R}}}\Bigg)nI_{\big(T_h > \frac{nS_{tcp}}{\mu}\big)}.
\label{equ:mtcp_app}
\end{equation}

\end{document}